\documentclass[10pt,twocolumn,usletter]{article}
\usepackage[top=2cm, bottom=2cm, left=2cm, right=2cm]{geometry}

\usepackage{amsmath,amssymb,amsfonts}
\usepackage{listings}                       %
\usepackage[noend]{algpseudocode}
\usepackage{multirow}
\usepackage{graphicx}
\usepackage{textcomp}
\usepackage{xcolor}
\usepackage[lined,boxed]{algorithm2e}

\usepackage{graphicx,subfigure}
\usepackage[bf]{caption}
\usepackage[sort,numbers]{natbib}
\usepackage{appendix}

\usepackage{xcolor}
\usepackage{booktabs}
\usepackage{graphicx}
\usepackage{enumitem}

\usepackage{times}

\usepackage[compact]{titlesec}
\titlespacing*{\section}{0pt}{*2}{5pt}
\titlespacing{\subsection}{0pt}{*2}{4pt}
\titlespacing{\subsubsection}{0pt}{*1}{1pt}

\definecolor{linkcol}{rgb}{0,0,0.5}
\definecolor{citecol}{rgb}{0,0.5,0.3}
\definecolor{urlcol}{rgb}{0.3,0,0}

\usepackage{xspace}

\let\OLDthebibliography\thebibliography
\renewcommand\thebibliography[1]{
  \OLDthebibliography{#1}
  \setlength{\parskip}{0pt}
  \setlength{\itemsep}{0pt plus 0.3ex}
}

\let\oldthebibliography\thebibliography
\renewcommand{\thebibliography}[1]{%
  \oldthebibliography{#1}
  \let\oldbibitem\bibitem
  \let\oldtextsc\textsc
  \def\oldbbland{et}
  \newcounter{authorcount}
  \def\bibitem[##1]##2{%
    \let\textsc\oldtextsc
    \let\bbland\oldbbland
    \oldbibitem[##1]{##2}%
    \let\textsc\mytextsc%
    \let\bbland\mybbland
    \setcounter{authorcount}{0}
  }
  \def\mybbland{\setcounter{authorcount}{0}\oldbbland}
  \def\dropetal##1.{ \bbletal}
  \def\mytextsc##1{%
    \oldtextsc{##1}%
    \stepcounter{authorcount}%
    \ifnum\value{authorcount}=5\relax%
      \expandafter\dropetal%
    \fi%
  }%
}

\usepackage[hang,flushmargin]{footmisc}

\renewcommand{\footnoterule}{%
  \kern -3pt
  \hrule width 1in
  \kern 2pt
}

\usepackage{url}
\makeatletter
\def\url@leostyle{%
  \@ifundefined{selectfont}{\def\UrlFont{}}%
  {\def\UrlFont{}}%
}
\makeatother
\usepackage[hyphenbreaks]{breakurl}

\definecolor{darkred}{RGB}{153,0,0}
\definecolor{darkblue}{RGB}{0,0,99}
\usepackage[colorlinks=true, linkcolor = darkred,   citecolor = darkred, urlcolor = darkblue]{hyperref}

\newcommand{\descr}[1]{\vspace{0.2cm}\noindent\textbf{#1}}
\newcommand{\reduce}{}

\usepackage{subfigure}

\newif\ifcomment
\commenttrue
\ifcomment
\newcommand{\jbnote}[1]{{\bf \textcolor{magenta}{JB: #1}}}
\newcommand{\edc}[1]{{\bf \textcolor{blue}{EDC: #1}}}
\newcommand{\alex}[1]{{\bf \textcolor{green}{AE: #1}}}
\else
\newcommand{\jbnote}[1]{}
\newcommand{\edc}[1]{}
\newcommand{\alex}[1]{}
\fi

\newif\ifhighlight
\highlightfalse
\ifhighlight

\else

\fi

\begin{document}

\sloppy 

\title{\bf ``Here's Your Evidence'': False Consensus in Public Twitter Discussions of COVID-19 Science\footnotemark}

\author{Alexandros Efstratiou$^1$, Marina Efstratiou$^2$, Satrio Yudhoatmojo$^3$,\\Jeremy Blackburn$^3$, and Emiliano De Cristofaro$^4$\\[0.5ex]
{\normalsize $^1$University College London, $^2$Independent Researcher, $^3$Binghamton University, $^4$UC Riverside}}

\date{}
\maketitle

\renewcommand*{\thefootnote}{\fnsymbol{footnote}}
\footnotetext{$^\star$Accepted for publication at 27th ACM Conference on Computer Supported Cooperative Work and Social Computing (ACM CSCW 2024). Please cite accordingly.}

\renewcommand*{\thefootnote}{\arabic{footnote}}
\setcounter{footnote}{0}

\begin{abstract}
The COVID-19 pandemic brought about an extraordinary rate of scientific papers on the topic that were discussed among the general public, although often in biased or misinformed ways.
In this paper, we present a mixed-methods analysis aimed at examining whether public discussions were commensurate with the scientific consensus on several COVID-19 issues.
We estimate scientific consensus based on samples of abstracts from preprint servers and compare against the volume of public discussions on Twitter mentioning these papers.
We find that anti-consensus posts and users, though overall less numerous than pro-consensus ones, are vastly over-represented on Twitter, thus producing a false consensus effect.
This transpires with favorable papers being disproportionately amplified, along with an influx of new anti-consensus user sign-ups.
Finally, our content analysis highlights that anti-consensus users misrepresent scientific findings or question scientists' integrity in their efforts to substantiate their claims.
\end{abstract}

\section{Introduction}\label{sec:intro}
As of April 2024, COVID-19 has claimed over 7 million lives worldwide~\cite{world_health_organization_who_2023}.
Despite the pandemic's severity, misinformation~\cite{al-rakhami_lies_2020,cha_prevalence_2021,kouzy_coronavirus_2020} and conspiracy theories~\cite{earnshaw_covid-19_2020,hughes_impact_2022,romer_conspiracy_2020} have circulated widely, often influencing real-world behavior~\cite{hughes_impact_2022,roozenbeek_susceptibility_2020}.
Although misinformation can be sourced from untrustworthy individuals or organizations~\cite{epstein_will_2020,yang_prevalence_2020}, even perceived experts sometimes use academic articles to promote anti-consensus views like opposition to vaccination~\cite{harris_perceived_2024}.
This can lead to misunderstandings about scientific support for official guidance since discussions of expert perspectives can be contentious.
For instance, after the Great Barrington Declaration~\cite{Lenzerm3908} %
suggested shifting away from lockdowns, the John Snow Memorandum %
countered it, advocating consensus on protective measures against COVID-19~\cite{alwan_scientific_2020}.

The pandemic's heavy politicization~\cite{green_elusive_2020,yang_prevalence_2020} further complicates the issue, as attitudes and behaviors %
differ among partisan groups~\cite{bruine_de_bruin_political_2020,havey_partisan_2020}.
Prior beliefs, values, and identities also influence acceptance of misinformation versus evidence-based claims~\cite{aghajari_reviewing_2023,efstratiou_adherence_2022}, impeding deliberation.
For example, people overwhelmingly target political opponents when correcting misinformation~\cite{allen_birds_2022}, but being corrected leads to elevated political bias in the content that a user shares~\cite{mosleh_perverse_2021}.
Such behavior can hinder effective responses to public health crises like COVID-19.

\descr{Motivation \& Research Questions.}
Some scholars project that, without interventions, unscientific views like the anti-vaccination movement could dominate %
within a decade~\cite{johnson_online_2020}, thus highlighting the importance of comprehending ``public science.''
To unravel the use of COVID-19 science in bolstering scientifically fringe beliefs, we incorporate insights regarding the role of social context and information-processing motivations beyond mere veracity~\cite{aghajari_reviewing_2023,buzzell_doing_2023,efstratiou_adherence_2022,kata_postmodern_2010,kata_anti-vaccine_2012}.
We propose that a false perception of consensus regarding anti-scientific views within the public can alleviate the psychological discomfort associated with holding these beliefs~\cite{efstratiou_adherence_2022}, thereby facilitating their propagation through misleading quixotisms~\cite{buzzell_doing_2023}, tactics, and tropes~\cite{kata_anti-vaccine_2012}.

More specifically, we identify and set out to address four main research questions:
\begin{itemize}[leftmargin=0pt]
	\item[] \textbf{RQ1.} Do public Twitter discussions of COVID-19 science follow scientific consensus? In other words, is the volume of public discussion of certain views commensurate to the volume of scientific papers supporting such views?
  \item[] \textbf{RQ2.} (a) Are certain sources disproportionately driving COVID-19 public conversations?
    (b) Is exposure to such sources associated with whether they espouse unscientific views?
  \item[] \textbf{RQ3.} How does COVID-19 scientific discourse evolve over time on Twitter?
  \item[] \textbf{RQ4.} In what other ways has COVID-19 science been (mis)represented in public circles?
\end{itemize}

\descr{Definitions \& Terminology.}
Our aim is to estimate scientific consensus and public consensus on several COVID-19 topics.
Through a comparison of scientific consensus and public consensus, we then estimate false consensus.
We define \textit{scientific consensus}, $C_s$, based on the volume of scientific papers promoting a particular standpoint, ${P_S}$, divided by the total volume of papers that either promote this standpoint or oppose it, $P_{\neg S}$, on a given issue, \textit{i}, i.e., $C_s(i) = \frac{P_S(i)}{P_S(i) + P_{\neg S}(i)}$.
For example, in the context of vaccines, scientific consensus would be the volume of papers opposing the efficacy or safety of COVID-19 vaccination over the total volume of papers that either support or oppose the efficacy or safety of COVID-19 vaccines (excluding papers on COVID-19 vaccines that do not promote a particular standpoint).

Similarly, we define \textit{public consensus}, $C_p$, as the volume of tweets disseminating scientific papers that promote a particular standpoint (e.g., those opposing vaccines, $T_S$) divided by the total volume of tweets disseminating scientific papers that either promote or oppose this standpoint ($T_{\neg S}$) on a given issue \textit{i}, such that $C_p(i) = \frac{T_S(i)}{T_S(i) + T_{\neg S}(i)}$.
Based on these metrics, we define \textit{false consensus}, $C_f$, as the discrepancy between scientific consensus and public consensus, i.e., $C_f = \frac{C_p}{C_s}$.
Henceforth, we refer to any papers that promote standpoints consistent with official guidance (e.g., those that support mask-wearing, social distancing, vaccinating, etc.) as \textit{conformist-consistent}, and those that promote standpoints inconsistent with this guidance as \textit{contrarian-consistent}.
Similarly, we refer to tweets and users promoting these standpoints and papers as \textit{conformist} and \textit{contrarian}, respectively. 

\descr{Methods.}
We source COVID-19 preprints from medRxiv and bioRxiv, and gather tweets mentioning them (or their peer-reviewed versions).
Using the retweet network, we classify users as either \textit{conformists} or \textit{contrarians} (Section~\ref{sec:classification}) and annotate a subset of preprints as \textit{conformist-} or \textit{contrarian-consistent}, estimating scientific consensus on several COVID-related topics and %
gauging false consensus (Section~\ref{sec:false_consensus}).
In particular, we examine whether the stances of papers or users correlate with Twitter attention (Section~\ref{sec:popularity}), and track tweet volumes and the emergence of contrarian and conformist accounts over time (Section~\ref{sec:temporal_analyses}).
Finally, we conduct a content analysis of selected cases where opposite-stance tweets drive Twitter engagement of papers, revealing tactics in COVID-19 science dissemination (Section~\ref{sec:case_studies}).

\descr{Main Findings.}
\begin{itemize}
\item We uncover a significant false consensus, with anti-consensus tweets being 2-10 times %
more prominent than the scientific evidence. %
\item Contrarian-consistent papers, although less numerous, are 69\% more likely to receive more Twitter mentions than conformist-consistent ones.
Also, contrarian users %
show increased activity over time, especially in later pandemic stages.
\item New contrarian account sign-ups increase after the pandemic's onset, with suspicious daily spikes.
In Twitter discourse, they sometimes misrepresent or cherry-pick findings from preprints, often against the authors' advice.
Also, they attempt to discredit research opposing contrarian views by suggesting corruption and conflicts of interest.
\item On the other hand, conformist users engage in more constructive scientific discussions, often exploring  details or sharing studies revealing nuances in COVID-19 measures.
\end{itemize}

\descr{Contributions.}
To the best of our knowledge, this mixed-methods study provides the first large-scale examination of COVID-19 science discussion in the public domain, along with an in-depth qualitative analysis of the motivations behind its dissemination.
Specifically, we provide the first quantification of COVID-19 scientific consensus, the public understanding of this consensus, and crucially, the juxtaposition between the two.
Alas, we find that scientific consensus is misrepresented on social platforms, causing a false impression of scientific division, driven by the disproportionate volume of secondary sources (users, tweets) over primary ones (scientific papers).
These misconceptions may normalize fringe viewpoints and bolster anti-institutional sentiment.
Moreover, we provide novel accounts of how this misrepresentation is carried out, through an increased volume of new contrarian accounts during the pandemic, as well as a sustained surge in activity of these accounts as the pandemic progresses.
Our work also situates established anti-science tactics, such as disingenuous attacks on scientists' integrity or selective discussion of narrative-affirming science~\cite{kata_anti-vaccine_2012}, to the COVID-19 context specifically.
These findings suggest that providing accurate consensus estimates to the public, especially during emerging crises like COVID-19, could help reduce confidence in misinformation and create more dissonance in movements that paradoxically weaponize science to promote anti-scientific views.

\descr{Ethical Statement.}
This study received ethical approval from the authors' institution (details are omitted for anonymity purposes).
We only collect and use the minimal amount of data required to perform our analyses as per our research questions and follow-up analyses, and avoid direct user quotations to protect their identities.

\descr{Disclaimer.}
Although we use the terms ``contrarian'' and ``conformist'' to generally refer to users that are against or for scientific consensus, respectively, we do not wish to imply the `correctness' of their respective views.
We only study how science is used \textit{holistically} among public circles and whether reasonable inferences are drawn from such discussions.
For example, it would not be wrong to say that COVID-19 vaccines may lose some efficacy against newer variants.
However, it is inaccurate to extrapolate this assertion to claims that COVID-19 vaccines are generally ineffective.
``Contrarians'' and ``conformists'' are merely a distinction based on how accurately they portray the state of scientific discourse.
We recognize that these terms carry inherent connotations; for example, COVID denialism or anti-establishment ideals may be some of the core narratives motivating the contrarian group.
However, we choose these labels due to the high-level nature of our work, as we are not surmizing that the same narratives are \textit{necessarily} motivating all participants.
Assuming that scientific consensus is a reasonable proxy for objective truth with respect to COVID-19, we are instead observing how Twitter discourse represents this truth; that is, whether it \textit{contradicts} it or \textit{conforms} to it.

Similarly, we are cautious not to describe scientific papers as contrarian or conformist; rather, we use the terms ``contrarian-\textit{consistent}'' and ``conformist-\textit{consistent}.''
In this way, we wish to convey that we classify papers based on whether their conclusions align with specific viewpoints and narratives, not on whether papers themselves have a certain ``agenda.''
Indeed, such a practice would be counter-intuitive.
Studying and documenting any risks associated with officially espoused measures, among other things, is crucial to forming robust, evidence-based policy.
Challenging conventional wisdom is pivotal in scientific advancement, and labeling science that brings new evidence to bear as ``contrarian'' would be a misnomer.
To this end, we also clarify that we are neither assuming nor implying any authors' stances or attitudes in classifying preprints.
Put simply, we are not raising any issues regarding how COVID-19 science is \textit{conducted}; this is beyond the scope of our work.
We are only studying how this science is \textit{disseminated} and \textit{discussed} to ensure that it has the intended impact.

\section{Background \& Related Work}\label{sec:background}

Though official health institutions are tasked with making evidence-based policy in consultation with several experts in the field who make up a consensus, trust towards such institutions is not always granted, and this consensus is rarely quantified or made perceptible to the public.
Thus, people may turn to sources who are in the minority but perceived as credible %
to substantiate contrarian viewpoints~\cite{efstratiou_misrepresenting_2021,harris_perceived_2024}.

The motivation behind this reliance is not immediately clear.
\citet{buzzell_doing_2023} posit an account of ``epistemic superheroism.''
Tracing its origins back to philosophers like Descartes and Kant, they argue that people often hold a romanticized view of knowledge production that emphasizes the individual rather than the collective.
Despite the increasing complexity of scientific disciplines making the attainment of all knowledge by a single individual impossible (and thus, we should rationally adopt views espoused by a collective of experts), people nonetheless embrace such ideas of ``epistemic heroics,'' leading to popular paralogisms such as ``do your own research.''

Other accounts attribute anti-scientific attitudes to postmodern ideas like the protest against institutional power and the rejection of epistemic certainty~\cite{kata_postmodern_2010}.
In this account, institutionalized science is viewed as corrupt and tyrannical, while ``enlightened'' scientists and doctors who fight ``the establishment'' break away and offer the people ``the truth''~\cite{prasad_anti-science_2022}.
This creates a paradox where, though one rejects expertise-based systems, they still rely on expertise to qualify their viewpoints when they perceive agreement between the two.
As~\citet{kata_anti-vaccine_2012} points out, the postmodern approach to medicine necessitates the use of several tactics to maintain a somewhat cohesive rationalization, such as misrepresenting scientific findings, attacking critics, and shifting hypotheses to alternative anti-establishment ideas when previous ones can no longer be entertained.
The result may be a skewed presentation of the extant evidence in public discussions, conflating selective exposure with the stance of the scientific establishment. %

\subsection{False Consensus and its Implications}

The public's perception of expert consensus influences health-related behaviors.
For instance, clear communication of doctors' support for COVID-19 vaccinations promotes vaccine uptake~\cite{bartos_communicating_2022}.
Contrarily, misunderstanding this consensus fuels vaccine hesitancy~\cite{motta_erroneous_2023} since misinformation is implied to originate from scientific sources~\cite{loomba_measuring_2021}.
Correcting misinformation on the Zika virus was more effective when conveyed by scientific authorities like the CDC~\cite{vraga_using_2017}.
However, in the highly-politicized context of COVID-19, this influence may be limited~\cite{helfers_differential_2023}, as entrenched opinions~\cite{aghajari_reviewing_2023} and anti-institutional sentiments~\cite{kata_postmodern_2010} may prevail.
Consequently, if people perceive false consensus, they may persist in espousing and acting upon misconceived beliefs.

False consensus, initially reported by \citet{ross_false_1977}, describes the tendency to overestimate the prevalence of one's own personality and beliefs.
This bias applies broadly, from climate change denial~\cite{leviston_your_2013} and 2020 US election denial~\cite{weinschenk_democratic_2021} to several other conspiracy beliefs~\cite{pennycook_overconfidently_2022}.
It is amplified among those with populist attitudes and hostile media perceptions~\cite{schulz_we_2020}.
Heavy social media users are particularly susceptible~\cite{bunker_how_2021}.
Also, engaging in extremist online communities can foster a skewed perception of extreme views' commonality~\cite{wojcieszak_false_2008,wojcieszak_deliberation_2011}.
Online echo chambers~\cite{cinelli_echo_2021,efstratiou_non-polar_2023,terren_echo_2021,zollo_debunking_2017} may worsen this effect, creating insular environments where individuals feel their views are widely shared~\cite{efstratiou_adherence_2022}. 

This may lead to something known as an illusion of consensus; people tend to give uniform weighting to claims with single or multiple primary sources as long as the number of secondary sources citing the primary ones are equal in number~\cite{yousif_illusion_2019}.
Whereas false consensus describes over-inflated prominence perceptions of certain views, illusory consensus outlines how the perceived credibility of different claims is affected by a fixation on secondary, instead of primary sources in support of them.
It may occur due to an implicit assumption that the primary sources are not truly independent~\cite{desai_getting_2022}.
Nevertheless, AI fact-checking agents are more effective in their corrections when citing multiple primary sources~\cite{ueno_trust_2023}, potentially indicating that illusory consensus is topic-specific or more prominent in human-to-human communications.

Illusory consensus has been documented in cases such as a majority of anti-climate change blogs using a single scientist with inadequate credentials to support their claims~\cite{harvey_internet_2018}, or minority perceived experts featuring heavily in anti-COVID vaccination Twitter clusters~\cite{efstratiou_misrepresenting_2021,harris_perceived_2024}.
In these instances, people potentially disregard scientific consensus in favor of who they view as righteous experts opposing the establishment~\cite{kata_postmodern_2010}.

\subsection{Misrepresentation of Science}
Amid the pandemic's onset, an overwhelming and conflicting influx of information potentially eroded trust in official institutions~\cite{zhang_shifting_2022}.
This climate may have fueled behaviors like the aforementioned epistemic heroics~\cite{buzzell_doing_2023} and postmodern reasoning~\cite{kata_postmodern_2010,kata_anti-vaccine_2012}.
However, such `lay science' poses concerns as it often disregards best practices, misinterprets data, and exhibits biases; for instance, misinformation tweets exploit retracted academic articles, sometimes politicizing retractions by invoking concepts like censorship~\cite{abhari_retracted_2023}.
Whereas retraction notices receive little attention compared to the original articles~\cite{serghiou_media_2021}.

Moreover, conclusions of e-prints may be misrepresented on platforms like Reddit~\cite{yudhoatmojo_understanding_2023}.
Some may engage in epistemic activism whereby they analyze data themselves to counter conclusions reached by official institutions, but committing errors like overlooking confounding factors in the process~\cite{lee_viral_2021}.
Beyond online social environments, recommendation systems on platforms like YouTube may influence exposure to pseudoscientific misinformation~\cite{papadamou_it_2022}.
Recently, \citet{beers_selective_2023} showed that Twitter users who oppose masks selectively share scientific work that aligns with their viewpoints while denouncing science that does not.

\begin{figure}
  \centering
  \includegraphics[width=0.65\columnwidth]{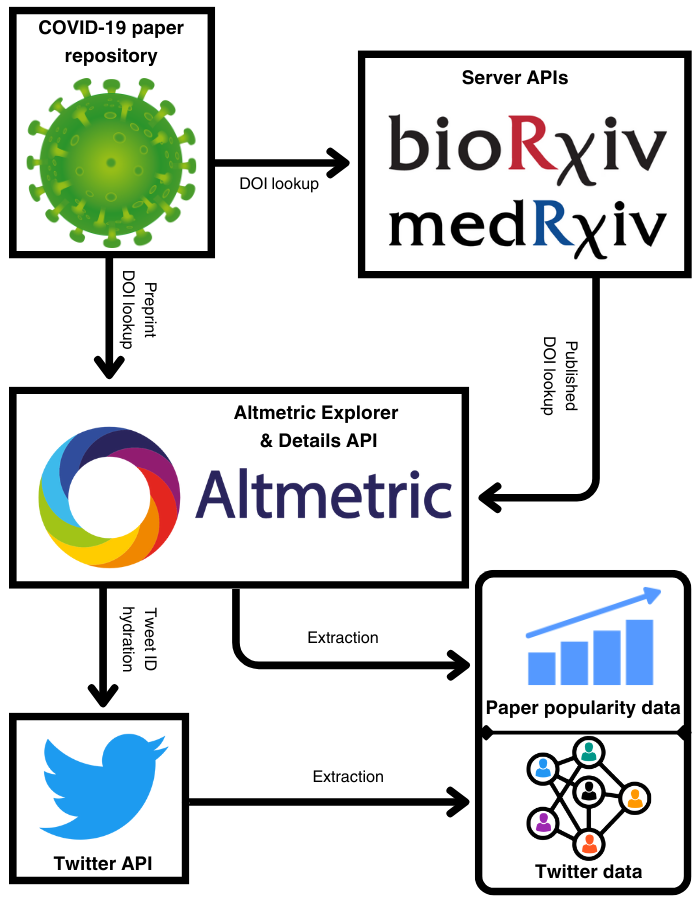}%
  \reduce
  \caption{Data collection pipeline.}
  \label{fig:data}
\end{figure}

\descr{Remarks.}
Scientific inquiry, though focused on accuracy and systematic analysis, may also serve personal sentiments~\cite{buzzell_doing_2023,kata_postmodern_2010,kata_anti-vaccine_2012} especially when enabled by social environments~\cite{aghajari_reviewing_2023}.
This can lead to false perceptions of consensus~\cite{desai_getting_2022,efstratiou_adherence_2022,efstratiou_misrepresenting_2021,harvey_internet_2018,yousif_illusion_2019} and influence behavior~\cite{hughes_impact_2022,roozenbeek_susceptibility_2020}.
This paper is the first effort to quantify false consensus in COVID-19 discussions and the extent to which public COVID-19 science representation is problematic.
We explore whether sources gain prominence based on the views they support, as well as how false consensus is propagated through a surge in contrarian activity and contrarian newcomers to the discourse.
Furthermore, we investigate tactics enabling scientific misinformation on COVID-19 and uncover that they resemble similar tactics that have been used for decades by traditional anti-science groups, such as the anti-vaccination movement~\cite{kata_anti-vaccine_2012}.
Arguably, our work informs science communication, health policy, and strategies to combat institutional distrust and polarization on scientific matters.

\section{Dataset}\label{sec:data}

As primary sources that contribute to scientific consensus, we collect all preprints from the joint medRxiv and bioRxiv COVID-19 repository\footnote{\url{https://connect.biorxiv.org/relate/content/}} released on or before November 4th, 2022.
Using Digital Object Identifiers (DOIs), we also query the medRxiv and bioRxiv APIs directly to obtain further information, e.g., whether they have been published in peer-reviewed venues or journals.

Then, we use the preprint and (where applicable) published DOIs to query the Altmetric Explorer and Details API.
Altmetric is a platform that measures the popularity of scientific papers, and its APIs provide us with metrics like the estimated number of Twitter mentions, news mentions, and total citations.
Through Altmetric Explorer, we also obtain IDs for any tweets that mention the corresponding DOI.
We hydrate these tweets using the Twitter API full-archive search v2 endpoint while also extracting tweets referenced by the initial set where applicable.
Our full data collection pipeline is shown in Figure~\ref{fig:data}.
We provide a description of our data in Table~\ref{tab:data}; note that the number of tweets and authors provided reflects only data used in our analyses (e.g., excluding non-English tweets, non-classified users, etc.)

\begin{table}[t]
  \centering
  \small
\setlength{\tabcolsep}{3pt}
  \begin{tabular}{lrrrrr}
    \toprule
\multicolumn{1}{c}{} & \textbf{Total} & \textbf{Published} & \textbf{Users} & \textbf{Retweets} & \textbf{Collection Date} \\
    \midrule
    {\bf Papers} & 25.3k & 13.4k & N/A & N/A & 04-Nov-2022 \\
    {\bf Tweets} & 1.24M & N/A & 346k & 70.8\% & 13-Dec-2022 \\
    \bottomrule
  \end{tabular}
  \caption{Overview of dataset.} %
  \label{tab:data}
\end{table}

\begin{figure*}[t]
  \centering
   \includegraphics[width=1.8\columnwidth]{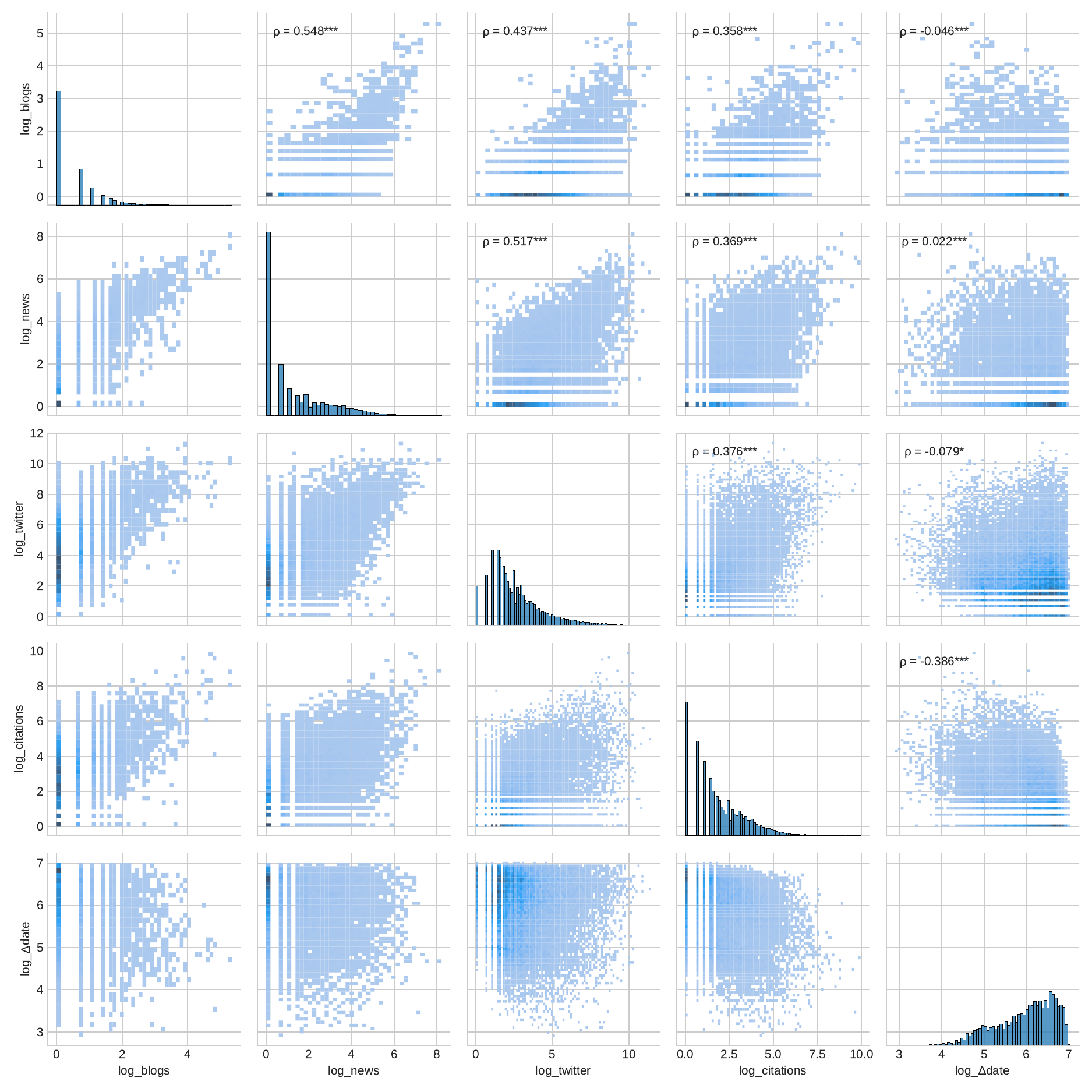}
  \reduce
  \caption{2D histograms of pairwise correlations between log-transformed metrics. ***\textit{p} $<$ 0.001, *\textit{p} $<$ 0.5.}
  \label{fig:pairplot}
\end{figure*}

We specifically focus on bioRxiv and medRxiv for two main reasons.
First, they provide a readily available repository specifically focused on COVID-19 preprints, allowing for a comprehensive collection of medical and biological COVID-19 science without the need for custom data collection pipelines and paper classification models.
Second, and most importantly, the medical and biological fields tend to produce research that directly informs health institutions' guidelines before recommending measures.
Thus, these fields are within the immediate scope of the present work, as it
concerns whether science that informs health authority guidelines is accurately represented within public circles.
We also recognize limitations regarding the use of preprints to reflect the state of scientific consensus en masse, although we focus on preprints because they are more accessible to the wider public, and to ward off the potential of publication bias in explaining our results.
Nonetheless, we reproduce our main analyses while restricting papers to those that eventually pass peer-review in Section~\ref{sec:illusion}, replicating our results.

As a preliminary exploration of the dataset, we analyze correlations between the papers' Altmetric popularity metrics.
We are especially interested in the relationship between citations and tweet counts, as this can indicate whether popularity among the general public might be associated with popularity within scientific communities.
Controlling for the effect of the number of news mentions, blog mentions, and the recency of the paper relative to the first COVID-19 preprint (lower value = older paper), we find a small but positive and significant Spearman's partial correlation between citations and number of tweets associated with a given paper ($\rho = 0.18, p > 0.001$).
We plot the pairwise correlations of the log-transformed metrics across both preprints and published versions in Figure~\ref{fig:pairplot}.

\begin{figure*}[t!]
  \centering
  \subfigure[Retweet network.]{%
  \label{fig:rt_network}%
  \includegraphics[width=0.325\textwidth]{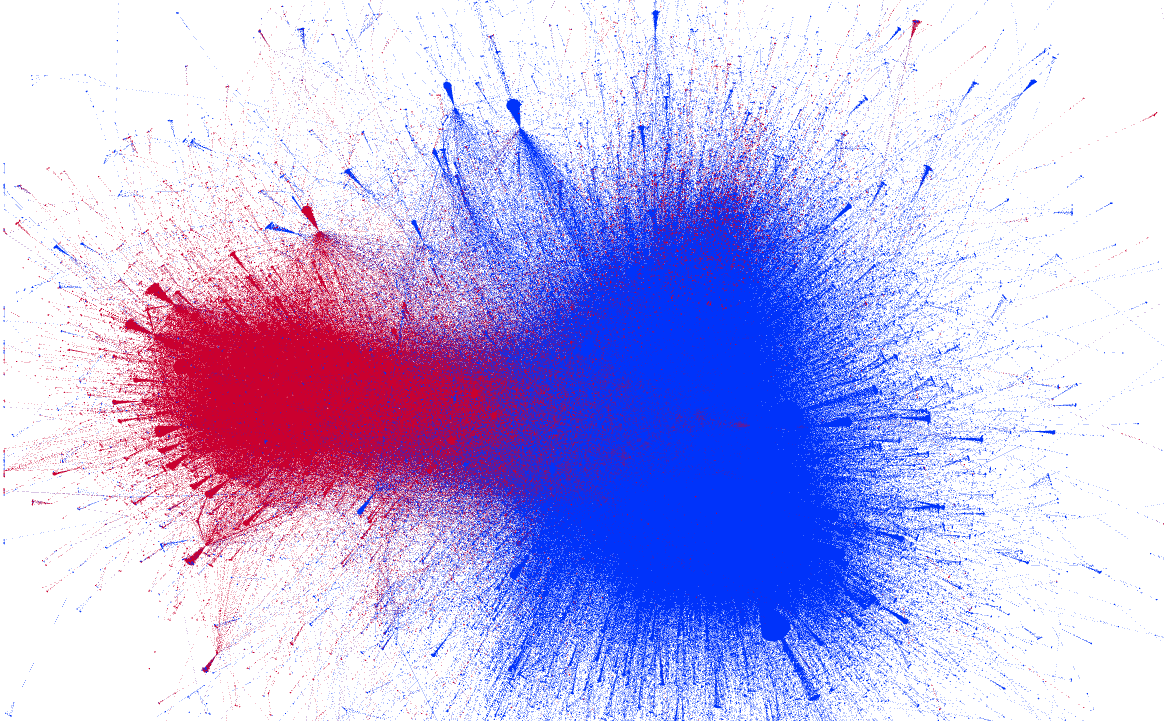}}%
  \hfill
  \subfigure[Quote tweet network.]{%
  \label{fig:quote_network}%
  \includegraphics[width=0.325\textwidth]{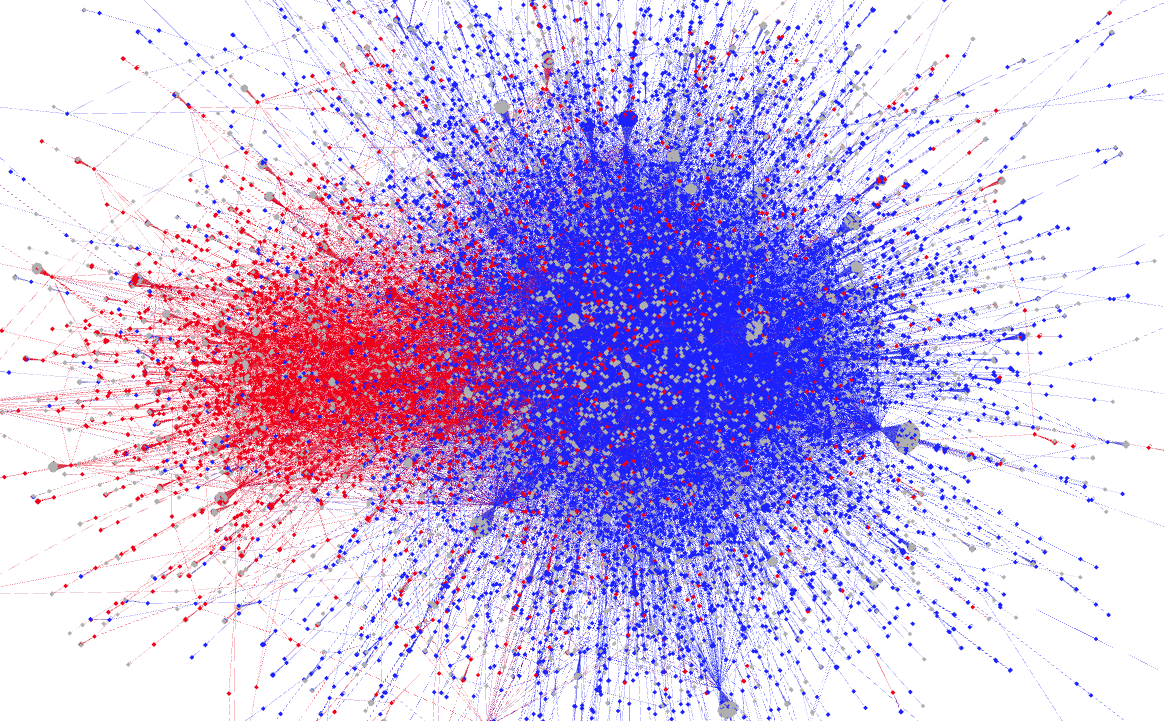}}%
  \hfill
  \subfigure[Reply network.]{%
  \label{fig:reply_network}%
  \includegraphics[width=0.325\textwidth]{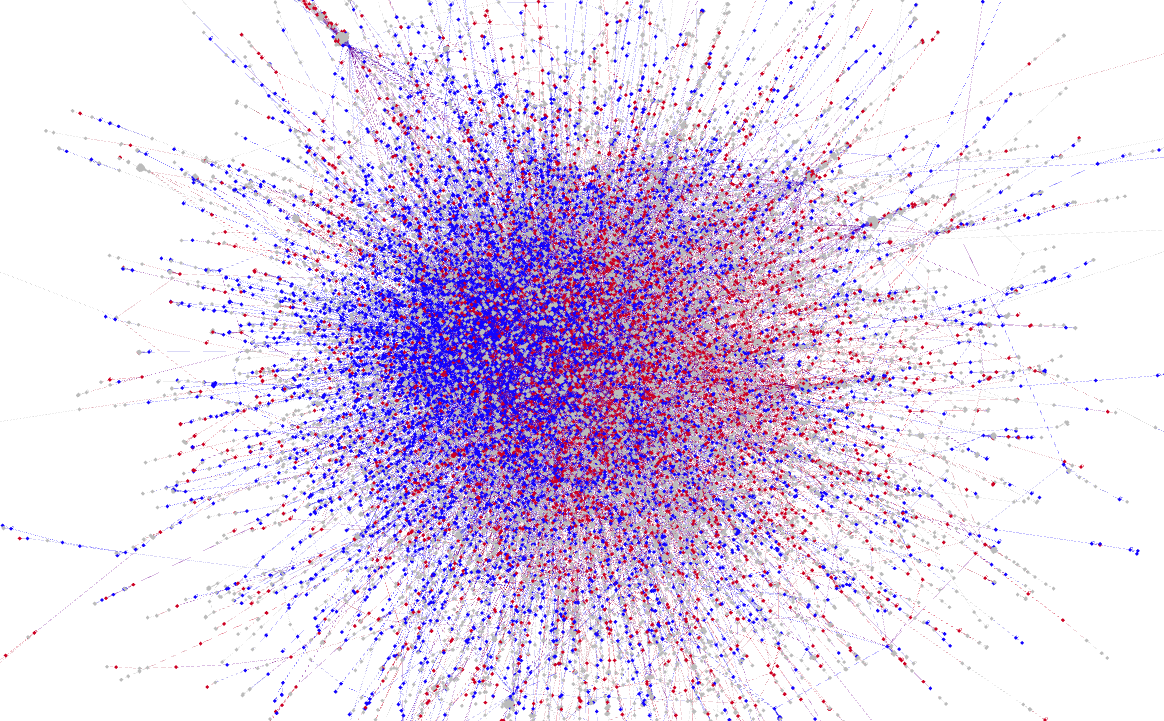}}%
  \reduce
\caption{Different types of networks colored by node label.}
\end{figure*}

Generally, although all correlations are statistically significant, the recency of a paper (i.e., $\Delta$date) only shows a noteworthy (negative) correlation with citation counts (i.e., older papers gather more citations).
This is potentially due to public or media mentions of scientific papers having a `spikier' exposure near their publication date, as opposed to citations which accumulate over a longer period.
All of the other popularity metrics (citations, tweets, news mentions, blog mentions) show small-to-moderate positive correlations between them.

\section{User Stance Classification}\label{sec:classification}
To enable us to address our four main research questions, we start by classifying users as pro- or anti-consensus.
This classification provides a baseline for comparing public perception against scientific consensus, tracking the evolution of COVID-19 science discussions, and contextualizing Twitter discourse.
As is typical with controversial topics~\cite{barbera_tweeting_2015,garimella_quantifying_2018}, we observe a segmented, two-community retweet network in preliminary visualizations.
Because retweets usually endorse the retweeted post~\cite{klein_attention_2022,metaxas_retweets_2014}, we use this to detect users' COVID-19 stance consistent with previous work on polarizing topics~\cite{klein_attention_2022,stewart_drawing_2017}.
We restrict the network to users who have posted at least one tweet in English.

\subsection{Node Classification}

Following the preliminary visualization, we treat this as a binary node classification problem.
The two possible classes are \textit{conformists} (i.e., those who follow official guidance) and \textit{contrarians} (i.e., those who oppose it).
We sample 200 random users with $\geq 3$ interactions in the retweet network.
Two authors annotate them based on their tweets' contents with a Kappa agreement of 0.61, which is considered substantial~\cite{landis_measurement_1977}.
Disagreements are resolved through discussion and a final agreement between the two annotators.
We do not observe cases where users share or express both contrarian and conformist views during this annotation exercise (e.g., users who are contrarian with respect to vaccines but conformist with respect to masks), suggesting that such cases are exceptionally rare.
Thus, we use the full retweet network instead of topic-specific ones for node classification.
Moreover, we are able to classify all users as either conformist or contrarian, thus we do not add an ``undefined'' class.

We extract the network's giant component and use the adjacency matrix and nodes' normalized spatial coordinates from a 2D ForceAtlas representation as training features.
We then train a 3-layer Graph Convolutional Network (GCN) with 5-fold validation, where 80\% of the data are used for training with the remaining 20\% used as a holdout test set at each fold, such that all of the data cycle through both training and testing phases.
Weights are reset prior to running every fold so that the model is fully blind to the test data.
We achieve an average accuracy of 98\% across all folds, suggesting a robust model that is not overfitting to the training data; the final model is trained on all 200 examples.

We elect this user-level network approach instead of a tweet-level language model because the two-community retweet network topology suggests the existence of two contested frames in Twitter discussions of COVID-19 science.
Several prior works have successfully utilized retweet networks in such contested frames contexts, like Black Lives Matter~\cite{klein_attention_2022,stewart_drawing_2017} or White Helmets discourse~\cite{wilson_cross-platform_2021}.
Although these works employ unsupervised, modularity-optimizing algorithms for community detection such as Leiden or Louvain, we use a GCN model for node classification for two reasons.
First, this allows us to set the number of potential classes a-priori informed by our own domain expertise, whereas unsupervised community detection methods can result in any number of communities.
Second, this approach enables us to manually annotate examples of users for training and testing based on their tweets' contents.
Therefore, despite the actual model relying on the retweet network itself to classify users, the model's accuracy is based on ground-truth assessments of these users' tweet contents, which increases our confidence in the model's performance.

We report the resulting graph in Figure~\ref{fig:rt_network}.
Out of 346k users, 101k (29.19\%) are contrarians.
We also visualize the quote (Figure~\ref{fig:quote_network}) and reply (Figure~\ref{fig:reply_network}) tweet networks; grey nodes reflect unclassified users who did not appear in the retweet network.
Otherwise, conformists are blue, contrarians are red.
The topological similarities between the retweet and quote networks indicate that users mostly add quotes supporting referenced tweets.
However, the reply network is fairly mixed, indicating discussions -- or debates -- between differently-opined users.

\subsection{Paper Prominence by Stance}

\begin{figure}[t!]
  \centering
  \includegraphics[width=0.8\columnwidth]{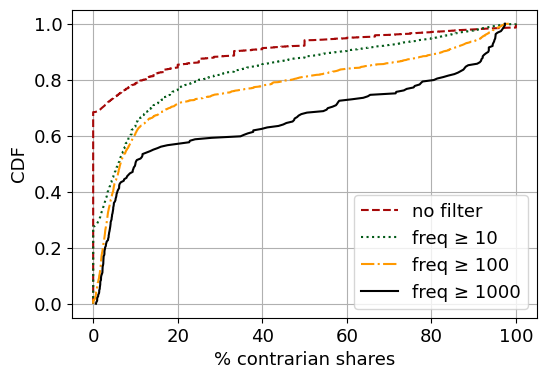}
    \reduce
  \caption{CDF of paper share distribution between conformists and contrarians.}
  \label{fig:papercdfs}
\end{figure}

Next, we examine the percentage of conformist or contrarian users' tweets mentioning each paper, such that papers with 0\% are shared solely by conformists and 100\% solely by contrarians. %
We plot the Cumulative Distribution Function (CDF) of these percentages in Figure~\ref{fig:papercdfs}, implementing tweet number filters to generalize patterns beyond papers with very few mentions. 

Most papers are predominantly shared by conformists (i.e., distribution $<$ 0.5) rather than following a bimodal distribution, but this diminishes at higher filtration magnitudes.
Thus, contrarians may rely more heavily on select popular papers. %
To confirm this, we produce two vectors that reflect the number of times contrarians or conformists mention each paper and compute their Gini coefficient (G).
G is a measure of inequality between values in a series, calculated as:

\begin{equation}
\small
  \centering
  G = \frac{\sum_{i=1}^{n}\sum_{j=1}^{n}|x_i - x_j|}{2n^2\bar{x}}
\end{equation}

Where $i$ and $j$ are index positions in vector $x$, and $n$ is the size of the vector.
G ranges between 0 and 1, and higher values indicate greater inequality, i.e., that a few papers concentrate more of the tweet mentions.
Indeed, we observe a slightly higher $G$ for the contrarian vector (0.89) than the conformist vector (0.85).
This pattern persists through the order-of-magnitude filters we implement above.

\subsection{Take-Aways}
Overall, we find ample presence of both contrarians and conformists in Twitter COVID-19 science discussions. The former rely on a smaller number of popular papers in their tweets, while the latter are more numerous.

\section{Quantifying False Consensus}\label{sec:false_consensus}
In this section, we use the distinction between contrarians and conformists to assess whether public discussions align with, or deviate from, scientific consensus (RQ1).
To determine a scientific consensus ratio that can be compared against tweet volume, we manually annotate random paper samples related to topics of interest.
We further use these annotations and user classifications to examine whether specific sources disproportionately drive COVID-19 conversations (RQ2a) and whether exposure to such sources is associated with the views they espouse (RQ2b).

\subsection{Topic Extraction}

\begin{figure}[t]
  \centering
  \includegraphics[width=0.99\columnwidth]{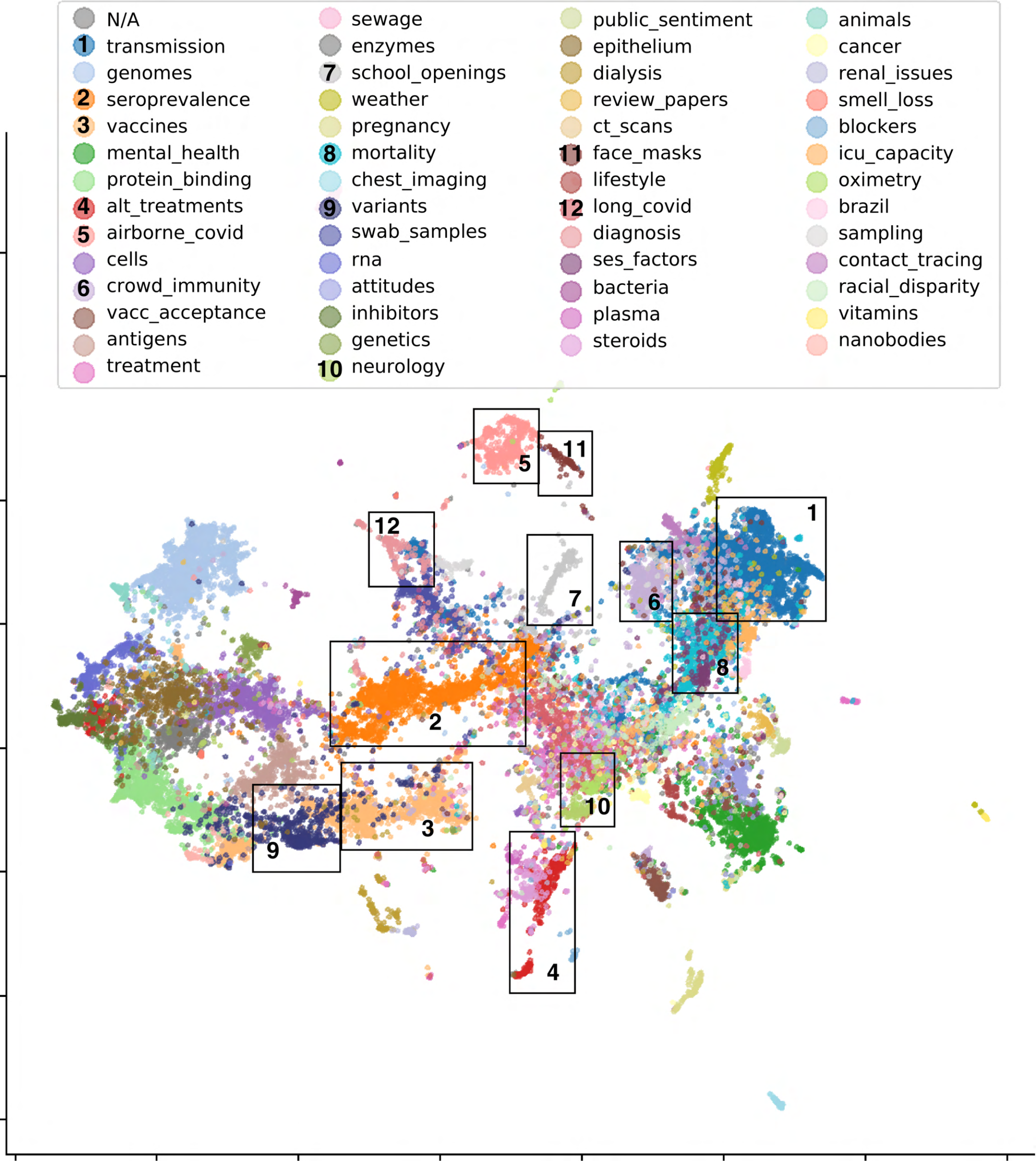}
    \reduce
  \caption{Topics and allocated papers shown as 2D UMAP embeddings. The first topic is a generic one for outliers. High-interest topics are numbered, and their respective positions are indicated in the plot.}%
  \label{fig:topics}
\end{figure}

We initially categorize papers into distinct topics by applying BERTopic~\cite{grootendorst_bertopic_2022}, which uses topic-specific word frequencies to increase granularity, on their titles and abstracts.
We extract a total of 54 topics reported in Figure~\ref{fig:topics} as 2D UMAP embeddings, and focus on the top 10 topics by frequency of tweets in two ways for further analysis.

First, we count the number of tweets mentioning the topic-specific papers.
The top 10 are: vaccines, variants (including vaccine effectiveness against them), genomes, seroprevalence (including fatality rates), transmission (i.e., lockdowns, asymptomatic transmission, social distancing, etc.), alt treatments (i.e., pharmaceutical drugs), mortality, long-COVID, enzymes (mostly binding behavior in vaccines), and airborne transmission (including use of face masks and surface vectors).
Overall, 65.6\% of tweets are about one of these 10 topics.

Second, we train another model on a sample of 100k tweets (excluding retweets).
This sample is substantial, as it amounts to just over one quarter of all eligible tweets (i.e., non-retweets), and we perform this sub-sampling due to memory constraints in the matrix representation of the training features.
We extract 27 topics, focusing again on the top 10.
The first two topics are generic outliers, likely due to Twitter's substantially shorter text.
From the remaining 8 topics, 3 refer to different facets of vaccines, while the others are on face masks, variants, virus origins, neurological damage, and airborne transmission.
Including the outliers, these top 10 topics account for nearly 90\% of the 100k tweets.

Using these insights, we arrive at the following categories to study further; these reflect compound themes made up of one or several high public-interest topics that we identify, and that account for the vast majority of public science discussions:

\begin{enumerate}
  \item {\em Vaccines.}
  This includes papers from the Vaccines and Crowd Immunity topics, as the latter mostly refers to immunity through vaccination or infection.
  It also includes papers from the Variants topic that assess vaccine efficacy against them (i.e., include the lemma ``vaccin'' in their titles).
  \item {\em Transmission and Non-Pharmaceutical Interventions.}
  This category captures lockdowns, face masks, whether COVID-19 is airborne, and other methods for curbing virus transmission.
  It is a composite topic of Transmission, Airborne Covid, School Openings, and Face Masks.
  \item {\em COVID-19 Dangers.}
  This refers to the risk of serious illness or death from COVID-19, and includes the following high-interest topics: Long Covid, Neurology, and papers from the Mortality and Seroprevalence topics which contain the lemma ``fatal,'' ``mortal,'' or ``death'' in their title (i.e., assessing fatality or mortality rates). 
  \item {\em Drug Treatments.}
  This only includes Alt Treatments, which concerns drugs that were either sanctioned (e.g., remdesevir) or not sanctioned (e.g., hydroxychloroquine) by official institutions.
  We use the World Health Organization's official recommendations to assess whether the treatment drug under study is recommended or not.\footnote{\url{https://www.who.int/publications/i/item/WHO-2019-nCoV-therapeutics-2022.4}}
  We filter papers to those that mention drugs for which the WHO has made a specific recommendation for or against in their title.
\end{enumerate}

\subsection{Illusion of Consensus}\label{sec:illusion}

For each category, we obtain 200 random preprint samples.
We then annotate their abstracts on whether they support or oppose official institutions' guidelines at the time of annotation, relying on category-unique codebooks (see auxiliary materials).
Broadly, for each category, we treat official institutions' standing as 1) supporting the use, efficacy, or safety of vaccines and boosters, 2) supporting the use of interventions and measures to thwart transmission, 3) suggesting that COVID-19 is a potentially dangerous illness, and 4) supporting the use of WHO-approved drugs and opposing the use of non-approved ones.
Annotations are performed by two authors, one of whom is a bioengineering expert, and disagreements are resolved through discussion to reach a final consensus (see Table~\ref{tab:consensus} for inter-rater agreements).

Scientific consensus ratios are estimated based on our annotations.
These ratios are then matched against the ratios of contrarian to conformist tweets (i.e., tweets from contrarian and conformist users, respectively) and contrarian to conformist users that tweet about a topic to determine the science-public discrepancy (i.e., false consensus).
We additionally annotate the 20 most mentioned papers per category to assess whether their ratios differ from the random samples.

\begin{table*}[t!]
  \centering
\setlength{\tabcolsep}{5pt}
  \small
  \begin{tabular}{lrrrrrr}
  \toprule
  \textbf{topic} & \textbf{$\kappa$} & \textbf{\%contrarian} & \textbf{\%contrarian} &
  \textbf{discrepancy} & \textbf{\%top20} & $\phi$ ($\phi_{authors}$) \\
  & & \textbf{papers} & \textbf{tweets (users)} &
  & \textbf{contrarian} &  \\
  \midrule
  vaccines & 0.66 & 4.2\% & 45.4\% (45.9\%) & 10.75 (10.86) & 50.0\% & 2.05 (2.07) \\
  NPIs & 0.59 & 6.5\% & 21.7\% (23.6\%) & 3.32 (3.61) & 31.3\% & 0.61 (0.69) \\
  dangers & 0.61 & 5.8\% & 29.6\% (32.8\%) & 5.10 (5.66) & 27.8\% & 1.02 (1.16) \\
  drugs & 0.63 & 23.8\% & 51.3\% (44.3\%) &  2.15 (1.86) & 68.4\% & 0.65 (0.48) \\
  \bottomrule
  \end{tabular}
  \caption{Agreement ratings ($\kappa$), percentage of contrarian-consistent papers, contrarian tweets/users, and contrarian-consistent top 20 most popular papers, along with the scientific-public discrepancy factor (i.e., \%contrarian tweets or users divided by \%contrarian-consistent papers.) Cramer's $\phi$ is a measure of the discrepancy's effect size based on $\chi^2$ tests (all significant at \textit{p} $<$ 0.001.)}
  \label{tab:consensus}
\end{table*}

Table~\ref{tab:consensus} shows significant and concerning gaps between scientific consensus and Twitter discussions across all four categories, with large effect sizes for all ($\phi > 0.5$).
These disparities hold even when only considering preprints that eventually pass peer-review.
Note that ratios exclude papers annotated as ``unclear'' (between 14\% and 29\% of the sample across categories), as per our definitions in Section~\ref{sec:intro}.

For drug treatments, contrarian discourse on Twitter is more than twice as prominent as expected based on scientific consensus.
This category exhibits the lowest discrepancy, possibly because several of the papers we classify as contrarian-consistent are not clinical trials and urge caution before making final recommendations in the drugs they study (see Section~\ref{subsec:limitations} for a related discussion).
In contrast, vaccine discussions on Twitter exhibit the largest discrepancy, with contrarian tweets being over ten times more prevalent than their consensus-derived ratio. This relative preference for contrarianism remains consistent at the user level.

\begin{figure}
  \centering
  \includegraphics[width=0.8\columnwidth]{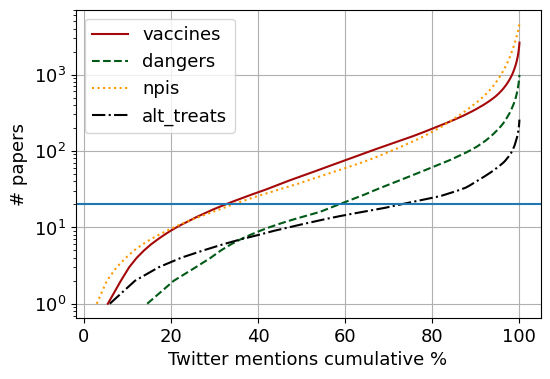}
  \caption{Cumulative percent of twitter mentions by number of papers per topic. The horizontal line denotes the 20th most mentioned paper in the category.}
    \reduce
  \label{fig:cumsum}
\end{figure}

The top 20 in each category are substantially more likely to be contrarian-consistent than the random sample, suggesting that highly-mentioned papers play a crucial role in driving false consensus.
For example, 50\% of the top-20 vaccine papers are contrarian-consistent, as opposed to just 4.2\% in the random sample.
Figure~\ref{fig:cumsum} shows that these papers amass between 32.6\% and 72.9\% of the total Twitter mentions for their respective topics.

\subsection{Engagement and Popularity Analyses}\label{sec:popularity}
Although we find some evidence that contrarian-consistent papers are more popular on Twitter, we set out to more formally quantify the degree to which this holds while also controlling for confounding factors.

Using Altmetric-estimated metrics, we regress the total tweet mentions of each paper on whether the paper is contrarian- or conformist-consistent.
We control for the number of authors, news mentions and citations, paper recency, topic, and whether the preprint has been peer-reviewed.
We also add citations and Twitter/news mentions from the published version of each paper if one exists.
As they are not random with respect to Twitter popularity, we omit the top-20 samples from this analysis.
The overall model is significant ($\textit{F}_{(9, 612)} = 90.81, \textit{p} < 0.001, adj. R^2 = 0.239$) and explains 23.9\% of the variance.
We show all coefficients and confidence estimates in Table~\ref{tab:mentions_regression}.

\begin{table}[t!]
  \centering
  \small
  \begin{tabular}{lrrrr}
  \toprule
  \textbf{variable} & \textbf{$\beta$} & \textbf{\textit{SE}} & \textbf{\textit{p}} & \textbf{95\% CI}\\
  \midrule
  intercept & -0.16** & 0.05 & 0.001 & [-0.25, -0.07] \\
  dangers\_topic & 0.03 & 0.07 & 0.65 & [-0.10, 0.16] \\
  npis\_topic & 0.06 & 0.08 & 0.48 & [-0.10, 0.22] \\
  vaccines\_topic & 0.10 & 0.12 & 0.37 & [-0.13, 0.34] \\
  contrarian & 0.57* & 0.24 & 0.015 & [0.11, 1.03] \\
  \#authors & -0.03 & 0.02 & 0.29 & [-0.07, 0.02] \\
  is\_published & 0.10 & 0.06 & 0.09 & [-0.02, 0.22] \\
  news\_mentions & 0.36* & 0.16 & 0.03 & [0.05, 0.67] \\
  citations & 0.12 & 0.10 & 0.20 & [-0.07, 0.32] \\
  $\Delta_{days}$ (recency) & 0.11** & 0.04 & 0.01 & [0.03, 0.19] \\
  \bottomrule
  \end{tabular}
  \caption{Full regression table for the number of paper Twitter mentions as the outcome variable. *\textit{p} $<$ 0.05, **\textit{p} $<$ 0.01. The drugs topic is used as the reference category when controlling for topics.}
  \label{tab:mentions_regression}
\end{table}

The number of news mentions, paper recency, and contrarian-consistent conclusions are significantly (and positively) associated with the number of Twitter mentions.
The lack of association with the number of citations/authors, publication status, and topic suggests that the public may not be as concerned with content or quality, instead anchoring on news availability and the view supported by the paper.
In fact, contrarianism shows the strongest effect out of all variables.

To corroborate this, we also compare contrarian- and conformist-consistent Twitter mentions using a Mann-Whitney U test, which makes no distribution assumptions.
This shows that contrarian-consistent papers (\textit{median} = 193) obtain significantly higher ranks than conformist-consistent ones (\textit{median} = 28), $\textit{U} = 33.0, \textit{p} < 0.001$, Common-Language Effect Size (\textit{CLES}) = 68.5.
That is, a randomly-chosen contrarian-consistent paper will accumulate more Twitter mentions than a randomly-chosen conformist-consistent one with a 68.5\% probability.
However, this advantage does not extend to contrarian \textit{users} or \textit{tweets} (see Appendix~\ref{app:twitter_eng}).
Moreover, we find no evidence that conformist-consistent papers are more likely to pass peer review than contrarian-consistent ones or vice-versa (Appendix~\ref{app:pub_odds}).

\subsection{Take-Aways}
RQ1 questions whether public Twitter discussions of COVID-19 science follow scientific consensus. 
Our analysis shows that they vastly deviate from scientific consensus as contrarian discourse occupies an incommensurate conversation volume.
In fact, this is possibly fueled by the disproportionate attention given to a minute fraction of papers, creating a false impression of scientific contestation.
Combined with our finding that a random contrarian-consistent paper has a 68.5\% probability of accumulating more mentions on Twitter than a random conformist-consistent one, we are able to answer RQ2a in that certain sources indeed disproportionately shape COVID-19 conversations.
This is also related to RQ2b on whether exposure to sources is associated with the views they espouse.
We find that contrarian sources enjoy more exposure when they are primary (i.e., papers) but not when secondary (i.e., tweets or users). 

\section{Temporal Analyses}\label{sec:temporal_analyses}

Having established relative imbalances in Twitter scientific discourse, we now study how this discourse evolves over time (i.e., RQ3).
Specifically, we analyze the volume of tweets about COVID-19 preprints over time, as well as whether this volume is driven by increased activity of existing users or the influx of new users to the platform.

\subsection{Tweet and Paper Volume}

\begin{figure}[t!]
  \centering
  \subfigure[Stacked plot of tweet volume by type.]{%
  \label{fig:ts_type}%
   \includegraphics[width=0.99\columnwidth]{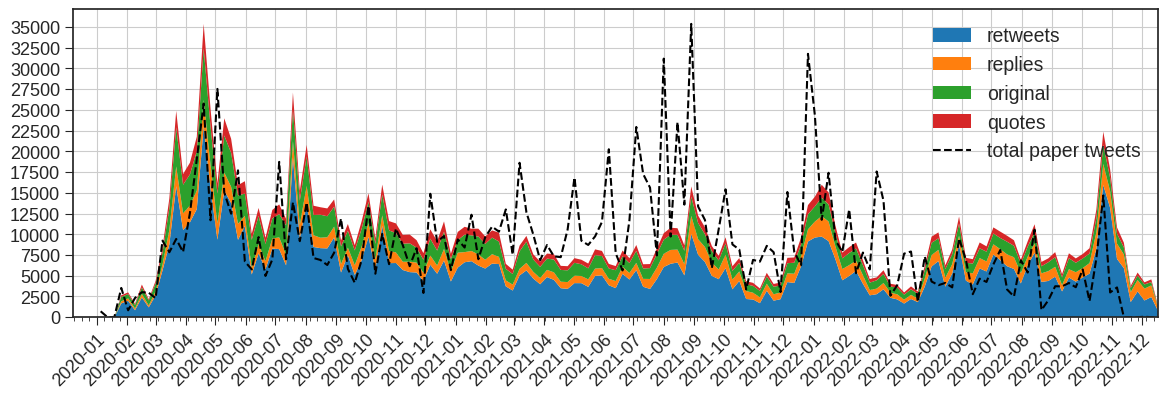}}%
  \qquad
  \subfigure[Unstacked area plot of tweet volume by group.]{%
  \label{fig:ts_group}%
   \includegraphics[width=0.99\columnwidth]{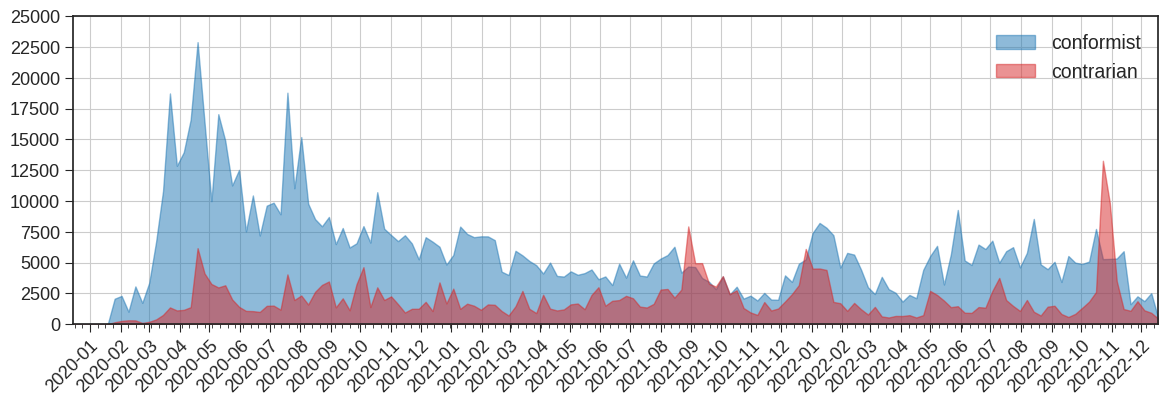}}%
  \qquad
  \subfigure[Stacked plot of volume of paper publications per week.]{%
  \label{fig:ts_pubs}
   \includegraphics[width=0.99\columnwidth]{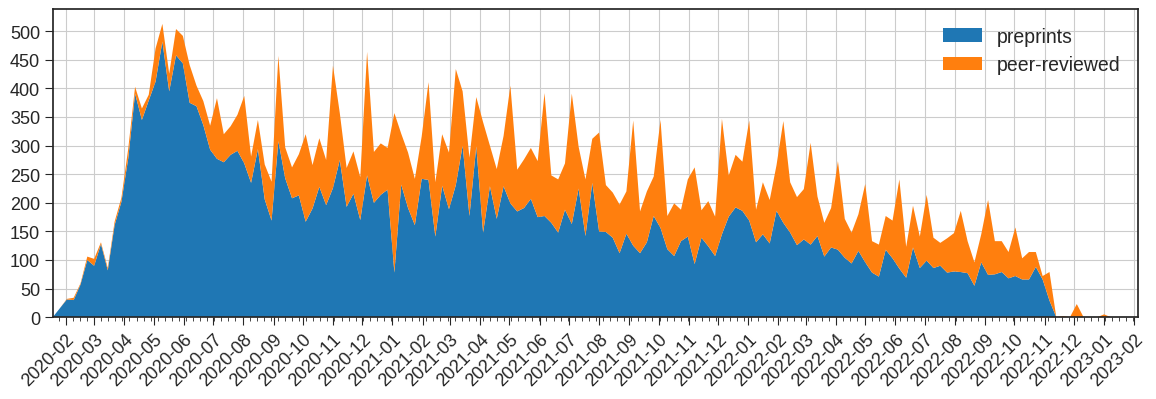}}%
  \reduce
\caption{Time-series plots of tweet volume. NB: Figure~\ref{fig:ts_type} includes a comparison to the Altmetric-estimated volume (dotted line).}
\end{figure}

We plot tweet volume time series, categorized by tweet type in Figure~\ref{fig:ts_type} and group in Figure~\ref{fig:ts_group}.
Activity spikes are mostly related to the release or publication of popular papers, as shown by the overlay of total Altmetric-estimated tweets for papers published in each period in Figure~\ref{fig:ts_type}.
Note that the overlay is scaled down to match tweet volume in our dataset.
Discrepancies in magnitudes may be due to our exclusion of non-English tweets, tweet removals from Twitter (but not from Altmetric estimations), and delays between paper publications and Twitter discussions.

The contrarian group was virtually non-existent until around mid-March 2020 when COVID-19 was officially declared as a pandemic by the WHO.\footnote{\url{https://www.who.int/news/item/29-06-2020-covidtimeline/}}
This is also when preprint publications spiked (Figure~\ref{fig:ts_pubs}), though Twitter discussions were already underway by then.
Overall, tweet volume does not closely follow publication volume, possibly because user attention is fixated on a small subset of papers.

\subsection{Timeline of Account Creations}

\begin{figure*}[t!]
  \centering
  \subfigure[Unstacked plot of monthly account creation volume by group since Twitter's inception.]{%
  \label{fig:ts_accts_ovrl}%
  \includegraphics[width=0.98\columnwidth]{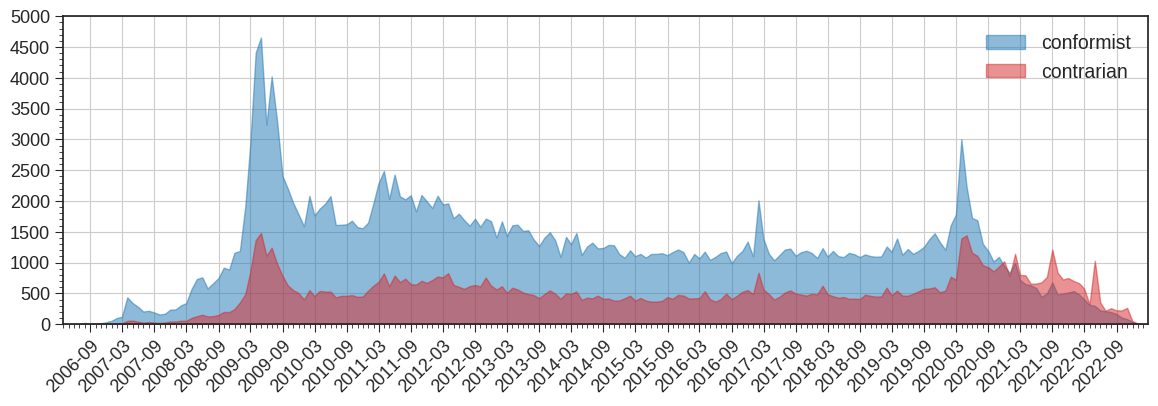}}%
  \qquad
  \subfigure[Unstacked plot of daily account creation volume by group during the COVID-19 period.]{%
  \label{fig:ts_accts_zoom}%
  \includegraphics[width=0.98\columnwidth]{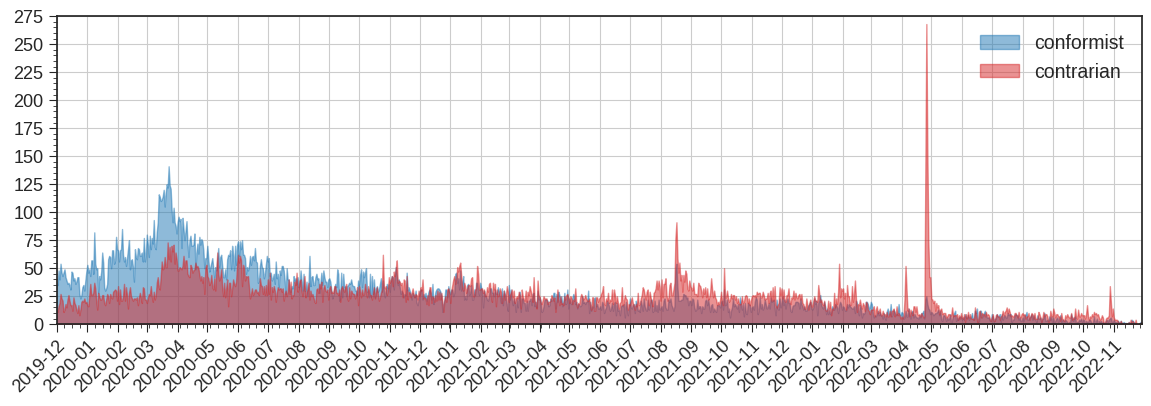}}%
    \reduce
\caption{Time-series plots of account creation volume.}
\end{figure*}

Next, we plot the volume of new account creations since the inception of Twitter (Figure~\ref{fig:ts_accts_ovrl}) and during the COVID-19 period, specifically, at a higher granularity (Figure~\ref{fig:ts_accts_zoom}).

From Figure~\ref{fig:ts_accts_ovrl}, we observe that conformist accounts were created at higher volumes throughout the years; contrarian accounts followed nearly identical patterns but in lower absolute numbers.
New Twitter sign-ups spiked for both groups around March 2020.
Contrarian account sign-ups overtook conformist ones towards the end of 2020.
This pattern continued until the end of our observation period.

Contrarian accounts also consistently overtake conformist accounts from the end of 2020 onward in terms of daily, not just monthly, sign-ups (Figure~\ref{fig:ts_accts_zoom}).
There is a suspicious spike in new contrarians between April 25-27, 2022, the largest for any 3-day period since Twitter's inception.
Accounts created in this period show elevated activity (along with other contrarians created in 2022), and seem to collaborate in amplifying already-prominent contrarian voices.
Please refer to Appendix~\ref{app:sus_accs} for more details.

\subsection{Take-Aways}

Our analysis sheds light on how COVID-19 scientific discourse evolves over time on Twitter (RQ3), in that it shows spikes near the declaration of the virus as a global pandemic, mostly owing to conformist users.
Thereafter, the discussion volume mostly follows the release of specific preprints or peer-reviewed papers.
Though conformist user accounts were historically created at consistently higher volumes than contrarian ones, this pattern switched approximately one year into the pandemic.
In addition, a suspicious influx of new contrarian sign-ups occurs in April 2022, which almost exclusively amplifies already influential contrarian users.

\section{Tactics and Motivations Behind Science Dissemination}\label{sec:case_studies}
In the previous sections, we performed aggregated, quantitative analyses of the false consensus phenomenon and its evolution.
Next, we set out to investigate whether science is also misrepresented on a \textit{qualitative} level (RQ4); e.g., whether certain papers are ``claimed'' by members of the public to support specific narratives.

As expected, contrarian- and conformist-consistent papers are mostly shared by contrarians and conformists, respectively.
In a Weighted Least Squares (WLS) regression, we confirm that our annotated labels significantly predict paper user distribution (a value from 0 to 1, with 1 meaning a paper is solely shared by contrarians) ($\beta = 0.655, \textit{SE} = 0.06, \textit{p} < 0.001$).
The overall model is significant and explains 65\% of the variance ($F_{1, 622} = 116.9, \textit{p} < 0.001, R^2 = 0.649$).
We weigh the variance contribution of each paper by its number of Twitter mentions. 

\begin{figure}[t!]
  \centering
  \includegraphics[width=0.99\columnwidth]{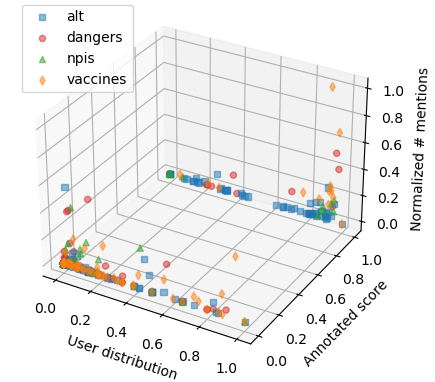}
\smallskip
  \caption{A 3D scatter plot of all annotated papers that received either a contrarian- (1) or conformist- (0) consistent classification, along with Twitter user mention distribution (1.0 meaning the paper was solely mentioned by contrarians) and (normalized) number of mentions.}
  \label{fig:3d_scatter}
\end{figure}

However, there are exceptions where user distributions do not align with annotations (outer-right and deep-left regions of the 3D scatter plot in Figure~\ref{fig:3d_scatter}).
As case studies, we focus on contrarian-consistent papers with user distributions of $\leq 0.40$ or conformist-consistent ones with $\geq 0.60$, garnering at least 100 Twitter mentions (12 total).
Thus, any time we refer to a ``case'' in the rest of this section, we mean one of these 12 preprints~\cite{hansen_vaccine_2021,buchan_effectiveness_2022,dorabawila_effectiveness_2022,cohn_breakthrough_2021,ioannidis_population-level_2020,arjun_prevalence_2022,rajasingham_hydroxychloroquine_2020,gautret_hydroxychloroquine_2020,lane_safety_2020,chen_efficacy_2020,garcia-beltran_multiple_2021,farinholt_transmission_2021}.

\subsection{Methodology}
In the rest of the section, we perform a qualitative content analysis of the tweets mentioning these papers.
Compared to narrative analysis (which focuses on individuals' personal stories about specific phenomena) or thematic analysis (which examines the \textit{meaning} behind qualitative data), content analysis is a higher-level method on the commonalities and patterns between data.
In other words, data is analyzed at face value.

For each case, we first familiarize ourselves with the paper in question and code tweets based on the claim(s) they make when citing it.
We focus on, at most, the top 200 most-retweeted unique posts. %
We then check the claims to gauge their accuracy %
and analyze the codes for emergent themes and commonalities.
These themes emerge from the juxtaposition of claims in tweets against claims in the papers 
and broadly reflect tactics or motivations behind the dissemination of COVID-19 science.

In the following, we do not directly quote users to protect their privacy but summarize and describe the most common talking points.
For each case, we first overview the paper's topic and main conclusions and then describe how it is discussed on Twitter.
Where cases were published in peer-reviewed journals by the time of data collection, we indicate this along with the publication journal when introducing them.
Otherwise, we do not mention publication status. 

\subsection{Theme 1: Selective Focus on Subset of Results}

For several papers, we observe a tendency for users to focus on a subset of results to strengthen their points while omitting other aspects of the paper, often misrepresenting the study's and authors' conclusions.

\descr{Vaccine efficacy against variants}.
Though cases in the remainder of this section are examined individually, we discuss Cases 1~\cite{hansen_vaccine_2021}, 2~\cite{buchan_effectiveness_2022} (later published in JAMA Network Open), 3~\cite{dorabawila_effectiveness_2022}, and 4~\cite{cohn_breakthrough_2021} for this theme in a ``batch,'' as they share several common characteristics---more precisely, they are all: 1) conformist-consistent but mostly discussed by contrarians, 2) vaccine-related, and 3) reporting waning vaccine efficacy against variants.
Cases 1-3 also report a peculiar ``negative effectiveness'' whereby vaccinations are associated with increased infection rates several months after immunization, though explicitly warning that this is possibly behavioral or statistical confounding (Case 2 eliminates this finding in a revised version which assesses \textit{symptomatic} infections specifically.) %
Case 4 assesses different types of vaccines and finds waning -- but still moderate -- protection for mRNA vaccines, with very low protection from a Janssen single-dose regiment against variants.
Crucially, all four studies report some benefit of vaccines, either against infection or severe disease, especially with booster doses (Cases 1-2).
Moreover, they explicitly support continued vaccination and boosting efforts, hence our conformist-consistent annotation.

The most-retweeted posts focus on the waning of vaccine effectiveness against variants and, in Cases 1-3, the ``negative effectiveness'' phenomenon that the papers' authors advise against taking at face value.
We also observe several users citing the studies to push against vaccine mandates (Cases 2-4) and supporting that vaccines do not work or are ineffective in the longer term (Cases 3-4).
Interestingly, we also find that the official account for the Sputnik V vaccine cites Cases 1 and 2 to suggest that its competitor mRNA vaccines are ineffective.

Some talking points are unique to individual studies.
For Case 1, users cite the study as a source for claims it does not make, for example, that natural immunity is better than vaccines or that vaccines damage the immune system.
Some use it to mock the vaccinated, while we also find a prominent user (who accumulates several retweets in the network) opposing a fact-check piece against one of their tweets that cites the study to make inaccurate claims.
The fact-check features one of the study's authors, who clarifies that interpretations other than the vaccine being effective are misrepresentative.

In Case 4, the most common talking point is an explicit focus on the low reported effectiveness of Janssen, with little mention of mRNA vaccines.
Some also argue that the constant need for boosting will render vaccines useless or that vaccines were inadequately tested before roll-out.
In the other cases, we find papers being used as sources to claim that unvaccinated people are supposedly vilified unnecessarily, to chastize official institutions like the CDC or FDA for supporting vaccines (Case 2), or that people who tried to bring these studies to light were censored (Case 3).

Among the few conformist posts we find, some focus on high reported vaccine effectiveness against severe disease (Cases 2 and 4) or interpret findings for policy (Case 3), where they argue for more evidence-based vaccination regiments and even critique the US government for using cheaper but reportedly less effective vaccines.
Two of Case 2's authors also appear in the conversation, clarifying their revised findings and re-iterating that boosters are effective.
Some conformists offer methodological critiques of studies (Cases 3 and 4), even arguing that the reported protection against severe disease, a conformist-consistent finding, may be unreliable and inconclusive (Case 4).
Other talking points for Case 4 include statements that waning efficacy against variants is unsurprising and a call for continued measures against COVID-19.

\descr{Relative vs. absolute mortality.}
Case 5~\cite{ioannidis_population-level_2020} (later published in Environmental Research) falls under the COVID-19 dangers topic.
It is contrarian-consistent but predominantly picked up by conformist users.
The paper attempts to estimate relative and absolute COVID-19 mortality risk across several European countries and US states, stratified by age group.
It reports that deaths for people under 65 years, especially those with no underlying health conditions, are ``remarkably uncommon.''
Its first version also reported generally higher relative risk in some US states compared to EU countries for people under 65.
However, these figures were updated with new data and more states in a revised version approximately one month later, which lowered relative risk estimates in US states.
In both versions, the authors advocate for shielding strategies that focus specifically on protecting the elderly or other high-risk individuals, hence our contrarian-consistent annotation.

The majority of tweets (80.9\%) are posted before the update, with 74.2\% being by conformists.
Most commonly, these focus on the elevated relative risk of under-65s in US states compared to EU countries, with some attributing this to healthcare inequality.
However, some posts criticize the paper.
One argument is that it is not sensible to estimate absolute mortality risk as early in the pandemic as the preprint's release, while another post invites those who dispute the paper's results to articulate their reasons for disagreeing. 

Discussions of this paper are more balanced, and there also are several contrarian posts.
Most commonly, these re-iterate the conclusions that under-65 deaths are uncommon and that high-risk populations should be prioritized, with arguments for re-opening the economy.
Occasionally, contrarians share the study's link and title without further context.
Due to the paper comparing COVID-19 mortality to driving-related fatalities, some users sarcastically support banning cars or call for a justification of why COVID-19 risks are not acceptable by authorities while driving risks are.

\descr{Vaccination and long-COVID odds.}
Also in the COVID-19 dangers topic, Case 6~\cite{arjun_prevalence_2022} examines long-COVID prevalence in hospitalized and outpatient COVID-19 cases. 
It reports a high long-COVID incidence, making it conformist-consistent with respect to COVID-19's potential danger. 
However, the study also reports that two vaccine doses predict higher long-COVID odds, which the authors describe as ``paradoxical.’’
Contrarians mostly highlight this, prompting one of the authors to add a comment on the preprint server providing evidence to the contrary and describing public discussions of their work as ``unfortunate ... cherry-picking’’.

As mentioned above, the most common talking point by contrarians focuses on the finding that vaccinated individuals have higher odds ratios of experiencing long-COVID. %
This is often presented as a stand-alone finding, though other users also mention some of the other predictors reported in the study (e.g., severity of COVID-19 or comorbidities).
Several tweets use the paper as a source to claim corruption in governments and the pharmaceutical industry, while some mock the authors' description of this result as paradoxical.

For the few conformist users who partake in discussions of the paper, most focus is on the relatively high prevalence of long-COVID in the sample.
Some also discuss the predictors of long-COVID as reported in the paper (including vaccination status.)

\descr{Remarks.}
Most cases of selective focus on a specific subset of results concern cherry-picking of potential anti-vaccine evidence by contrarian users, usually against the advice or conclusions of the original papers' authors.
We also find one case where conformist users selectively focus on relative risks of COVID-19 death, although the paper they cite concludes that absolute risks are low.
Occasionally, conformist users focus on some methodological details of papers, even criticizing findings that would generally align with a conformist point of view.
However, a focus on results is more common.
Nonetheless, conformists tend to discuss results holistically, e.g., supporting more efficacious vaccines over less efficacious ones. 
On the other hand, contrarians mostly take selected findings at face value to support narratives such as corruption or incompetence of official institutions. 

\subsection{Theme 2: Discrediting and Diminishing}

In two cases, we find that Twitter users share studies that do not align with their viewpoints to refute their findings or conclusions, either by questioning the authors' integrity (contrarians) or the study's methodological robustness (conformists).

\descr{Insinuations of financial conflicts.}
Case 7~\cite{rajasingham_hydroxychloroquine_2020} (later published in Clinical Infectious Diseases) examines HCQ as a COVID-19 prophylactic and finds no significant reduction in COVID-19 incidence.
Although categorized as conformist-consistent, it mostly attracts contrarians for two reasons.
First, it reports a positive but statistically non-significant effect of HCQ.
Second, it acknowledges funding from the Minnesota Chinese Chamber of Commerce (MCCC), which contrarians use to question the study despite a clarification that funders played no role in conducting the study.

By far, the most common talking point is about how the MCCC purportedly paid the authors to discredit HCQ.
This is the case for the most-retweeted post, and also with an account responsible for one-third of all posts for this paper.
The account seems automated and cites alleged corruption in replies that it directs in an untargeted manner, often even replying to irrelevant political posts.
Other posts argue that the positive effect shown in the paper could turn significant if pooled with other studies in a meta-analysis, thus using this study as evidence \textit{for} HCQ.
The few conformists who post about this paper focus on the non-significant effect of HCQ, with some suggesting that the study should end the debate around the drug.
We also find one case where conformists, much like contrarians, focus on the reported positive effect of HCQ instead of statistical significance.

\descr{Discussions of methodological limitations.}
Case 8~\cite{gautret_hydroxychloroquine_2020} (later published in the International Journal of Antimicrobial Agents), classified as contrarian-consistent, finds that HCQ is effective against COVID-19, especially if used alongside azithromycin.
Many think that it has initiated the debate around HCQ for COVID-19, as Donald Trump referred to it while still serving as US president~\cite{wong_hydroxychloroquine_2020}.

We find this study primarily discussed by conformists.
Several posts discuss its methodological limitations or amplify others that do; these include small sample size, non-randomization, non-blinding, and that authors disregard patients who end up in intensive care or die.
Others urge caution in over-relying on a single study and call for more evidence on the drug, while a few highlight that HCQ may be unsafe for use.
Some express cautious optimism that the paper's results can be replicated in larger studies.

For contrarians, most users simply disseminate the study.
Occasionally, they also share the study as pushback to claims that HCQ does not work against COVID-19, often several months after the release of the preprint.
One user cites the study to lambast social media companies that purportedly censor those who claim HCQ's effectiveness.

\descr{Remarks.}
In two instances, users mention studies to disagree with them explicitly.
Contrarians use one study's funding source to question the results and authors' integrity, though the authors state that the funders had no input in the study.
Conformists mostly dispute one study's findings on methodological grounds without showing any strong sentiment on the study's subject matter (HCQ) at the time; some even express optimism about its results.

\subsection{Theme 3: Focus on Contextual Factors}

In Case 9~\cite{lane_safety_2020} (later published in The Lancet Rheumatology), conformist users draw attention to a contrarian-consistent paper while focusing on contextual factors surrounding it.
The study analyzes HCQ's safety and reports short-term safety when prescribed alone but elevated cardiovascular risks when prescribed with azithromycin.
The paper does not ``position'' itself as contrarian per se, but we assign a contrarian-consistent label due to the HCQ safety finding.

Users either focus on the dangers of the HCQ-azithromycin combination or convey the study's findings.
In response to the FDA's Emergency Use Authorization for the drug combination at the time, some tweets also criticize the decision to prescribe the drugs before enough evidence was available.
Some of these are directed at then-US president Donald Trump, chastising him for praising the drug combination.

Of the few contrarians that discuss the study, most circulate it without further context, while others focus on the short-term safety of HCQ alone.

\subsection{Theme 4: Dissemination of Oppositional Findings}

Occasionally, users share papers whose findings do not generally align with the group's standing but nonetheless provide more information on COVID-19 treatments or vaccines.
We observe this behavior exclusively for conformists.
Specifically, we find one case where conformists share promising early findings about using HCQ (before a wider consensus on the drug emerged) and two cases of papers suggesting that virus variants may escape COVID-19 vaccines.
Again, none of the papers position themselves as contrarian per se, but we assign contrarian-consistent labels due to potential talking points associated with them (e.g., ``vaccines do not work against variants'').
Other papers under Theme 1 also find waning efficacy against vaccines but are classified as conformist-consistent.
This is because those papers indicate that vaccines continue to have \textit{some} effectiveness against variants and/or recommend continued vaccination efforts in the abstract, as per our codebook.

\descr{Hope in pre-consensus evidence.}
Case 10~\cite{chen_efficacy_2020}, released in April 2020 before a broader consensus emerged, shows that HCQ reduces recovery time and promotes pneumonia absorption.
It is mostly discussed by conformists with mixed reception.
Some tweets highlight study limitations like sample size and lack of clarity behind methodological choices, with some even insinuating that its conclusions are unreliable.
Others express hope that HCQ may be a promising drug in treating COVID-19, though such expressions are reserved, often with calls for further evidence.

Contrarians either simply disseminate the study, or focus on the positive HCQ effect found in the paper, sometimes as a response to claims that HCQ does not work.
One contrarian user who remains critical of HCQ throughout the pandemic also condemns the study's methods.

\descr{Clarifying scientific implications.}
Case 11~\cite{garcia-beltran_multiple_2021} (later published in Cell) concludes that variants may escape mRNA vaccine neutralization.
Conformist users mostly circulate or summarize the study and discuss the concerning findings around immunization escape. 
One conformist science journalist suggests that the paper raises deep concerns about vaccines but receives criticism from others for being unduly ``alarmist.''
One of the authors also responds to the journalist, urging caution in interpreting results as the study is an in-vitro examination of a pseudovirus, not a direct assessment of vaccine effectiveness.
Generally, conformists do not criticize the study and instead share it for informational purposes.

The few contrarians mentioning the study either highlight the reduction in neutralizing activity against variants or suggest that vaccines are not adequately protective.

\descr{Information-sharing and urging vigilance.}
Case 12~\cite{farinholt_transmission_2021} (later published in BMC Medicine) studies six vaccinated people infected with COVID-19 and concludes that the Delta variant may possess vaccine evasion properties.
It is predominantly discussed by conformists on Twitter.

Most highlight the conclusion that Delta may escape vaccination.
The paper is also used as a source to stress Delta's danger and the importance of protecting oneself, that transmission can occur in several places (the case study investigated an outdoor setting), and as a call for additional measures such as boosters.
Others draw distinctions between the protection afforded by different types of vaccines.

Among the few contrarians, one cites the paper to suggest that vaccines are ineffective and alternative treatments (presumably HCQ) should not be dismissed.
Some state that Delta can evade vaccines, while others highlight that the vaccinated can still transmit the virus, show symptomatic illness, or die.
Some also mock vaccine effectiveness.

\descr{Remarks.}
Overall, conformist users often disseminate papers that may complicate some of their bottom lines (e.g., that vaccines are effective), but they do so for informational purposes to highlight some intricacies around public response to COVID-19.
For the two vaccine-related papers, conformist users either highlight some potential early warnings (Case 11) or the need for continued vigilance against new variants that may escape vaccine immunity (Case 12).
In the case of the HCQ-related paper, when the HCQ debate was still ongoing, conformist users disseminated early findings on the potential effectiveness of the drug and expressed prudent optimism for future studies.

\subsection{Take-Aways}

Our qualitative analysis provides a more nuanced picture of how COVID-19 science is discussed on Twitter.
First, some of the tactics we uncover mirror those reported in previous work~\cite{kata_anti-vaccine_2012,yudhoatmojo_understanding_2023}, e.g., cherry-picking results or alluding to corruption to question the accuracy of findings that do not align with certain viewpoints.
We observe this predominantly for contrarian users.
Second, contrarians tend to discuss (often misrepresented) take-home points of different studies, whereas conformist users tend to also touch on other subtleties of academic papers, such as methodological limitations and problematic conclusions.
Third, conformists are willing to circulate data regardless of the conclusions they arrive at.

Overall, the conformist group may engage in more good-faith scientific discussions even in the public sphere, whereas the contrarian group's engagement with COVID-19 science seems more performative and ideologically motivated.
To answer RQ4, misrepresenting scientific findings on Twitter is not the norm, but when it does happen, such misrepresentations can garner significant traction.
Occasionally, biased tactics like cherry-picking and bad-faith accusations of scientific dishonesty are employed.
However, aside from some motivated discussions predominantly among contrarian circles, there are also somewhat nuanced scientific discussions in the public sphere on the part of the conformist group.
Thus, public platforms like Twitter can act both as hubs of COVID-19 (mis)information and as communities to engage with emerging science.

\section{Discussion and Conclusion}

This paper examined COVID-19 science discussions on Twitter, showing that public discussions of COVID-19 science result in vast misrepresentations of scientific consensus across several topics.
Scientific papers are often cited in a biased manner to support specific viewpoints instead of engaging in good-faith debates, resulting in preferential treatment of certain sources over others.
Making scientific consensus more directly perceptible by the public can help diminish the extent of this false consensus, which can, in turn, incentivize better policy-making.

Our work identified and addressed four main research questions.
First, we assessed whether Twitter discussions align with scientific evidence ({\bf RQ1}); we uncovered significant and consistent false consensus effects across several COVID-19 issues.
That is, viewpoints with limited scientific backing appear disproportionately prominent on Twitter.
We also focused on exposure to COVID-19 science sources ({\bf RQ2}) and found that a few papers incommensurately influence public discussions (RQ2a).
Contrarian-consistent sources gain more exposure than conformist-consistent counterparts when they are primary (i.e., scientific papers) but not secondary (i.e., tweets or users; RQ2b).

Next, we studied the temporal evolution of this conversation, activity, and new accounts around the declaration of a global pandemic (\textbf{RQ3}).
Contrarian users became more active during the pandemic, both in terms of tweet volume and new account creation.
Moreover, we found a collaborative effort to amplify prominent contrarian voices, even though we could not verify if this activity was organic or coordinated.
Finally, we examined whether/how COVID-19 papers are misrepresented (\textbf{RQ4}).
While not the norm, misinformation about the contents of studies occasionally surfaces and garners attention.
Some users, particularly contrarians, selectively present findings to bolster their views or question opposing research.
Nevertheless, conformist users often engage in nuanced methodological discussions and share findings beneficial for the COVID-19 response, regardless of alignment with their stance.

\subsection{Implications}

COVID-19 misinformation has been rampant throughout the pandemic~\cite{madraki_characterizing_2021,wang_understanding_2023}, often employing the moral values and emotions of people who encounter it to assist its spread~\cite{solovev_moral_2022}.
However, little work focused on perceived experts as potential sources of this misinformation~\cite{efstratiou_misrepresenting_2021,harris_perceived_2024}, and, to the best of our knowledge, no other studies quantified the discrepancy between scientific consensus and {\em perceived} consensus on social media.

The public was repeatedly advised to follow the science or listen to the experts, yet the severe false consensus we find could mean that it is unclear what the science says or who the experts are.
Combined with elevated distrust in institutions~\cite{adhikari_trust_2022,caplanova_institutional_2021,kata_postmodern_2010,yuan_different_2022} that act as aggregators of current scientific opinion, it is likely that some people may approach this problem with arbitrary criteria such as satisfying existing views~\cite{aghajari_reviewing_2023}, placing themselves in emergent groups~\cite{efstratiou_adherence_2022}, or valuing conclusions that they have reached themselves over those reached by collective efforts~\cite{buzzell_doing_2023}.
Meanwhile, the mistaken perception that scientific opinion is somewhat split can make adherence to COVID-19 misinformation more comfortable~\cite{efstratiou_adherence_2022}.

We also find that engagement with COVID-19 science does not always occur in good faith.
Rather, users may adopt the arbitrary criteria we refer to above, thus using science to argue for a particular position instead of enriching their knowledge.
This is a major concern since science itself may be weaponized and disseminated in a combative way, potentially opening an opportunity for malicious actors to use it for their own ends~\cite{oreskes_merchants_2011}.
In turn, under this guise of ostensible credibility, harmful information can come to be adopted and more widely disseminated by genuine users~\cite{starbird_disinformation_2019}.
Indeed, we observe some suspicious patterns in new account creations and automated activity in pushing back against undesirable findings.
Moreover, scientific preprints are also used to promote interests, as in the case of the Sputnik V account attempting to cast doubts on the safety and effectiveness of its competitor mRNA vaccines.

Effectively, although scientific consensus on topics such as COVID-19 is measurable and quantifiable, it is not directly perceivable by the wider public.
Rather, appeals to consensus rely on converging opinions of official institutions, as was the case with anthropogenic climate change~\cite{oreskes_scientific_2004}.
However, as mentioned earlier, trust in these institutions is variable.
It was not until \citet{cook_quantifying_2013}'s landmark study that a more digestible figure of scientific consensus emerged, reporting that 97.1\% of published research on the topic endorsed the idea of anthropogenic climate change.

Here, we similarly demonstrate that the overwhelming majority of COVID-19 scientific literature is aligned with the positions of official institutions.
Nonetheless, the balance of public conversations does not reflect this consensus.
Instead, some users engage in performative discussions where science is used as a tool for motivated reasoning~\cite{taber_motivated_2006}, eschewing the pursuit of epistemic truth~\cite{kata_postmodern_2010}.

Public sentiment should ideally reflect scientific consensus as closely as possible for topics where this consensus is the best proxy for objective truth (e.g., COVID-19 and climate change).
False consensus allows duplicitous doubting of the evidence~\cite{oreskes_merchants_2011} and, consequently, the entertainment or even adoption of empirically flawed policies. %
For example, the very misperceptions we demonstrate have been used as reasoning for anti-scientific and potentially harmful policy positions by Florida's Governor and Surgeon General~\cite{bendix_desantis_2023}.
Our findings highlight the need to quantify true scientific consensus on topics for which public and policy opinion deviate from the extant evidence.

This is an open challenge for HCI and CSCW communities, as existing CSCW frameworks can lend themselves to addressing this problem in practice.
Misinformation, both about COVID-19 specifically (e.g.,~\cite{bozarth_wisdom_2023,wang_understanding_2023}) and on general issues more broadly (e.g.,~\cite{drolsbach_diffusion_2023,efstratiou_adherence_2022,hussein_measuring_2020,mcclure_haughey_misinformation_2020,moran_sending_2023}), has received a lot of attention within the CSCW community especially with respect to how it is situated between social and technological systems.
Due to challenges like lower trust in institutions that attempt to curtail problematic content~\cite{bozarth_wisdom_2023} and the rapid rate at which misinformation proliferates, there have been recent calls to adopt approaches that target the social norms that enable the spread of misinformation, rather than individual pieces of information themselves~\cite{aghajari_reviewing_2023}.
Indeed, convergence towards such perspectives arises from long-standing work in the field that treats information as latent \textit{narratives}~\cite{nied_alternative_2017,wilson_cross-platform_2021,wilson_assembling_2018}.
Adherence to misinformation may then be understood as deep-rooted motivations, such as the need to protect one's social identity or the need to resolve ambiguity around complex events~\cite{efstratiou_adherence_2022}, which are served by these narratives.

The idea that science around COVID-19 does not align with official guidelines is, itself, such a narrative.
Crucially, belief in this narrative could then give rise to more layered beliefs, such as the idea that powerful institutions like social media platforms are censoring the truth~\cite{wilson_cross-platform_2021}; something which our findings refute, since contrarian viewpoints do not reflect the extant evidence, nor are they censored based on the prominence of these views on Twitter.
Successfully challenging these narratives to minimize reactance is something that can be informed by design perspectives within CSCW.
For example, some works have proposed how subtle nudges may be used to address cognitive biases during information parsing like targeting the balance of availability between misinformation and corrections~\cite{konstantinou_nudging_2023}.
Other researchers have put forward recommendations for designing tools in science communication specifically, like providing analytics that enable science communicators to assess the audiences they reach and the impact of their efforts~\cite{williams_hci_2022}.
These design insights may be adapted to make scientific consensus more easily perceptible by the public, ranging from dynamic estimations of scientific consensus as it evolves, to implementing methods of digital delivery on educating users and broader networks on this consensus.
The combination of misinformation and design frameworks can enable CSCW scholars to more effectively challenge complex and deep-seated latent beliefs, potentially introducing dissonance in people who adhere to these narratives and assisting them in forming more accurate perceptions~\cite{efstratiou_adherence_2022}.

Reliance on official institutions alone as messengers may be insufficient for two reasons.
First, trust in these institutions cannot always be assumed~\cite{kata_anti-vaccine_2012}.
Second, in their role as aggregators of consensus, institutions may inadvertently reverse \textit{perceptions} of such consensus.
Our study highlights how many secondary sources cite a small number of primary sources to create an illusion of consensus; this can, in turn, give credibility to ideas without sufficient evidence to support them~\cite{yousif_illusion_2019}.
With the institutional approach, the opposite occurs; a large number of primary sources are narrowed through the lens of only a handful of secondary sources that are not personified and, thus, perhaps not as accountable in the eyes of the public.

Having said that, we do not argue that institutions should avoid taking an explicit position on these issues.
Rather, the \textit{basis} of these positions should be directly perceptible since calls to follow the science or listen to the experts can be misconstrued in biased ways.
Although we do not anticipate this will completely eliminate the paradoxical use of science to support unscientific views among the public, we hope it will highlight false consensus fallacies and, by extension, the divergence from what bodies of expertise support.
Once again, how perceptible and available these scientific bases are may depend on how complete the computer-supported pipelines for uncovering them and relaying them to the public are.

\subsection{Limitations}\label{subsec:limitations}

Naturally, our work is not without limitations.
Although our abstract-count approach is similar to prior work~\cite{cook_quantifying_2013}, we acknowledge that paradigm shifts or scientific opinion switches can occur owing to authoritatively robust and large studies.
Indeed, particularly with respect to recommended pharmaceuticals for the treatment of COVID-19, we find that some official guidelines changed throughout our observation period.
We do not consider these shifts in annotating papers (i.e., we only consider official guidelines at the time of annotation), since papers that could be considered consistent with official guidelines at the time may continue to be mentioned on Twitter after the guidelines change.
However, we recognize limitations with this approach, as this may inflate the percentage of papers that we annotate as contrarian-consistent.
For example, some papers put forward preliminary findings supporting the use of HCQ which was under emergency use authorization for the treatment of COVID-19 by the US Food and Drug Administration (FDA) until June 15th, 2020, when this authorization was revoked in the face of new evidence.\footnote{\url{https://bit.ly/3QH6R1C}}
To that end, another potential method of estimating scientific consensus would be a direct survey of experts who have published work in a given field or do active research within it.
Nonetheless, we have no reason to believe that our main conclusions would change had we followed this approach, especially since we were fairly conservative in developing our codebooks.
On the contrary, given findings like early hope for HCQ as a treatment drug followed by subsequent convergence against it, we expect that direct surveying of experts would only amplify our results.
Overall, our case studies reveal that the authors of the studies we annotate tend to agree with our labels; cases where this is not as obvious were labeled as contrarian-consistent, which further supports that the false consensus effect we find may be even more pronounced in reality.

Furthermore, we do not appraise methodological robustness or confidence in each study's findings.
However, we stress that we only aim to study \textit{false consensus} and \textit{illusory consensus}; that is, the deviation between the \textit{number} of primary and secondary sources behind certain claims, and how that may affect perceived norms on Twitter.
This is also why we explicitly study tweets that directly link to these primary sources.
Quality control would be best situated within the scope of studies that appraise epistemic certainty itself, for example, meta-analyses or systematic reviews.
Regardless, our cursory observations and quality-proxy analyses (e.g., publication odds) suggest that quality appraisal would not alter our main conclusions. 

Finally, the false consensus effect we report is derived from Twitter posts.
This does not necessarily mean that the percentages we report reflect the opinions or attitudes of the entirety of Twitter, let alone the general population.
For example, misinformed users may simply be more likely to share posts than others~\cite{pennycook_fighting_2020}; without a fully random sample, it is difficult to estimate how well this false consensus effect generalizes. %
We also note that, while the behavior of some accounts suggests automated or at least inorganic activity, we do not eliminate these accounts from our analyses for two major reasons.
First, due to changes to the platform's API, we were unable to verify whether these accounts are indeed coordinated.
Second, these accounts can act as information brokers for other organic users, such that removing them would obfuscate the nature of COVID-19 science discussions and their spread.
Still, our findings reflect what may be directly perceivable on social platforms like Twitter, which means that opinions and behaviors may still be unduly influenced owing to descriptive social norms~\cite{cialdini_descriptive_2007}.

\descr{Acknowledgements.} This material is based upon work supported by the National Science Foundation under Grant No.~IIS-2046590, the UK's EPSRC grant EP/S022503/1, which supports the UCL Centre for Doctoral Training in Cybersecurity, and the UK's National Research Centre on Privacy, Harm Reduction, and Adversarial Influence Online (REPHRAIN, UKRI grant: EP/V011189/1).
Any opinions, findings, and conclusions or recommendations expressed in this work are those of the authors and do not necessarily reflect the views of the NSF, UKRI, or EPSRC.

{\small
\bibliographystyle{myabbrvnat}

}

\appendix

\section{Twitter Engagement Analyses}\label{app:twitter_eng}

Beyond Twitter engagement at the paper level, we also assess whether a user's or tweet's stance is associated with Twitter engagement.
We compute the total retweets that each user accumulates on their COVID-19 paper tweets and regress this on user stance while controlling for number of followers and accounts they are following, total tweets posted by the user, number of tweets that appear in our dataset, user verification status, and account age (in days).
The model is significant and explains a moderate degree of variance ($\textit{F}_{(7, 81529)} = 18.25, \textit{p} < 0.001, R^2 = 0.379$), but only the number of followers significantly predicts total retweets accumulated ($\beta = 0.62, \textit{p} = 0.02$).
User stance does not show significant effects ($\beta = -0.003, \textit{p} = 0.55$).
All continuous variables are z-score standardized, and we employ robust standard errors.

At the (English-language) tweet level, however, we observe a slightly different pattern.
We regress the retweets obtained by individual tweets on whether a contrarian or conformist user posts the tweet.
We control for all of the account features used in the user-level analysis while additionally controlling for the age of the tweet itself, and whether the tweet is a reply, a quote tweet, or original; again, we employ z-score standardization and robust standard errors.
This results in a significant overall model ($\textit{F}_{(9, 360621)} = 149.5, \textit{p} < 0.001, R^2 = 0.118$), though only 11.8\% of the variance in the outcome variable is explained.
We show the full regression in Table~\ref{tab:rts_regression}.

\begin{table}[t!]
  \centering
  \small
  \begin{tabular}{lrrrr}
  \toprule
  \textbf{variable} & \textbf{$\beta$} & \textbf{\textit{SE}} & \textbf{\textit{p}} & \textbf{95\% CI}\\
  \midrule
  intercept & 0.007** & 0.002 & 0.009 & [0.002, 0.011] \\
  quote & -0.044*** & 0.002 & $<$ 0.001 & [-0.049, -0.039] \\
  reply & -0.058*** & 0.002 & $<$ 0.001 & [-0.063, -0.054] \\
  acct\_age & -0.0009 & 0.002 & 0.59 & [-0.004, 0.002] \\
  tweet\_age & -0.002 & 0.002 & 0.34 & [-0.005, 0.002] \\
  followers & 0.33*** & 0.030 & $<$ 0.001 & [0.267, 0.385] \\
  following & 0.04** & 0.016 & 0.007 & [0.012, 0.074] \\
  num\_tweets & -0.022*** & 0.004 & $<$ 0.001 & [-0.029, -0.015] \\
  verified & 0.135*** & 0.011 & $<$ 0.001 & [0.114, 0.156] \\
  contrarian & 0.021*** & 0.002 & $<$ 0.001 & [0.017, 0.025] \\
  \bottomrule
  \end{tabular}
  \caption{Full regression table for tweet-level retweets as outcome variable.}
  \label{tab:rts_regression}
\end{table}

All variables except account and tweet age show statistically significant effects; however, the largest effect is exerted by the number of followers, followed by whether the tweeting account is verified.
Tweets posted by contrarian users are also significantly associated with more retweets; however, the magnitude of this effect is negligible.

\section{Paper Publication Odds}\label{app:pub_odds}

We conduct a multiple logistic regression to assess whether a paper's stance (contrarian- or conformist-consistent) influences its likelihood of passing peer review while controlling for the number of authors, paper recency (difference in days from first preprint in the sample), and topic (dummy-coded on drug treatments).
We include all annotated papers not classified as unclear, resulting in \textit{N} = 689.

The model is statistically significant (\textit{p} $< 0.001$), but the overall fit is low (pseudo-$R^2 = 0.03$).
As expected, the number of authors ($\beta = 0.35, \textit{p} < 0.01$) and paper recency ($\beta = -0.39, \textit{p} < 0.001$) show significant and moderate effects.
Paper stance does not statistically influence publication odds ($\beta = -0.20, \textit{p} = 0.40$), though we note the non-negligible effect size.
Thus, owing to the relatively low number of contrarian-consistent papers, this finding may be inconclusive due to potentially limited statistical power in detecting a true effect.

\section{Suspicious accounts}\label{app:sus_accs}

We focus some analyses on the 268 accounts created in the suspicious spike between April 25-27th, 2022.

\subsection{Suspicious Account Activity}

We compare suspicious account activity to randomly sampled contrarian accounts.
Using a 1000-iteration Mann-Whitney bootstrapping procedure, we find that suspicious accounts post significantly more tweets per day (\textit{median} = 21.75) compared to the control group (\textit{median} = 5.50. \textit{SE} = 0.02) with a large effect size (\textit{CLES} = 72.24, \textit{SE} = 0.03, \textit{p} $ < 0.001$).
We observe similar results when comparing the suspicious accounts to the entire rest of the dataset (\textit{median} = 5.48, \textit{CLES} = 72.25, \textit{p} $< 0.001$).
However, these findings do not replicate when comparing suspicious accounts to contrarian accounts created specifically in 2022.

\begin{figure}[t!]
  \centering
  \includegraphics[width=0.99\columnwidth]{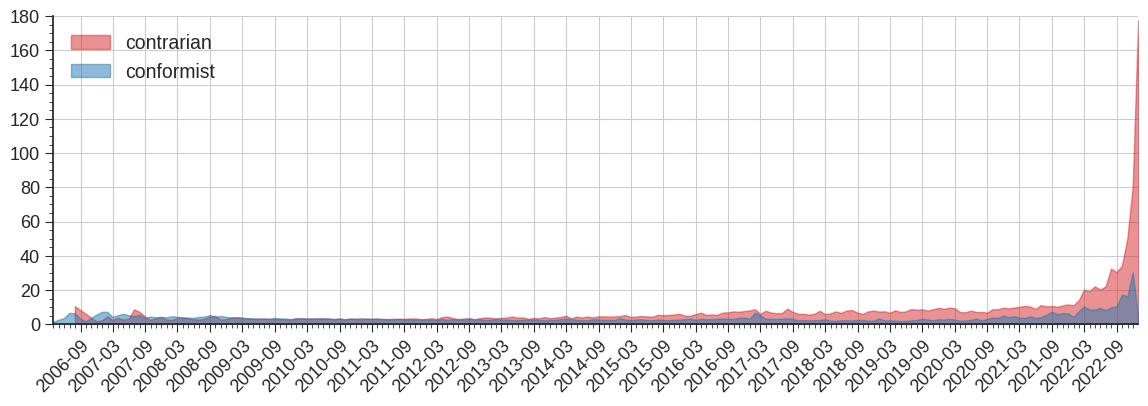}
  \vspace{-0.2cm}
  \caption{Median tweets per day by account creation date per group.}
  \label{fig:ts_activity}
\end{figure}

When comparing suspicious accounts to conformist accounts (\textit{median} = 3.06, \textit{SE} = 0.01), the effect is even more pronounced (\textit{CLES} = 77.43, \textit{SE} = 0.03), indicating that high activity is a characteristic of recently-created contrarian accounts.
We also replicate this finding when comparing them to conformist accounts created specifically in 2022 (\textit{median} = 9.41), albeit with a lower effect size (\textit{CLES} = 62.41, \textit{p} $< 0.001$).
This suggests that unusually high activity is a trait of recently created contrarian accounts, as depicted in Figure~\ref{fig:ts_activity}.

It is unclear whether this is due to new accounts generally being more active close to their creation, if more recent accounts are lesser-known and therefore unmoderated, or if something else drives this pattern.
Unfortunately, due to changes in the Twitter API while this project was ongoing, we are unable to collect further data to shed more light on this.

\subsection{Suspicious Account Amplification}

\begin{figure*}[t!]
  \centering
  \subfigure[Lorenz curves of out-degrees per group and period.]{%
  \label{fig:lorenz_out}%
   \includegraphics[width=0.80\columnwidth]{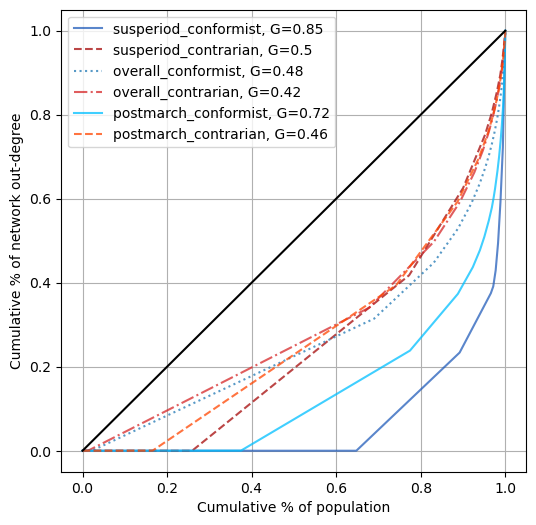}}%
  \qquad
  \subfigure[Lorenz curves of in-degrees per group and period.]{%
  \label{fig:lorenz_in}%
   \includegraphics[width=0.80\columnwidth]{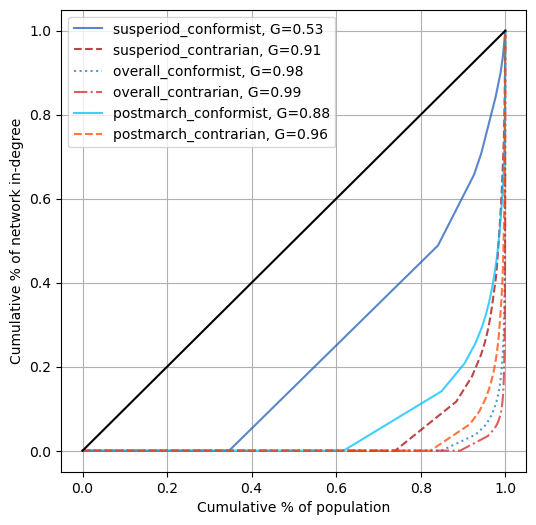}}%
\caption{Lorenz curve plots of network degree inequality.}
\end{figure*}

We study whether these accounts disproportionately amplify specific users in the scientific COVID-19 discourse.
We obtain two separate directed retweet networks for retweets by accounts created in the April 25-27 spike (i.e., \textit{susperiod}, short for suspicious period) and the general period from March 2022 onward (i.e., \textit{postmarch}, including susperiod), and compare them to the overall network in Figure~\ref{fig:rt_network}.
Then, we extract node out-degrees (i.e., number of retweets they post) and in-degrees (i.e., number of times they are retweeted) for conformists and contrarians, which we use to plot Lorenz equality curves along with Gini coefficient calculations in Figures~\ref{fig:lorenz_out} and~\ref{fig:lorenz_in}, respectively.

The black diagonal represents perfect equality.
Steeper Lorenz curves imply higher Gini coefficients, indicating greater inequality where a few nodes accumulate most degrees.
Figure~\ref{fig:lorenz_out} shows conformists have more out-degree inequality, especially in restricted periods.
Contrarians exhibit greater in-degree inequality, particularly during these periods.
Thus, contrarians are more consistent in how many tweets they repost but only amplify a few accounts.
From a cursory examination, the amplified accounts are already prominent voices in the overall period.

\end{document}